\pdfoutput=1
\RequirePackage{ifpdf}
\ifpdf % We~are running pdfTeX in pdf mode
\documentclass[pdftex]{sigma}
\else
\documentclass{sigma}
\fi

\numberwithin{equation}{section}

\newtheorem*{Theorem*}{Theorem}

 { \theoremstyle{definition}

 }

\usepackage{mathrsfs}

\def\pa{\partial}

\newcommand{\tr}{\operatorname{Tr}}

\def\one{\hbox{{1}\kern-.25em\hbox{l}}}

\begin{document}
%\allowdisplaybreaks

\renewcommand{\thefootnote}{}

\newcommand{\arXivNumber}{2301.12848}

\renewcommand{\PaperNumber}{042}

\FirstPageHeading

\ShortArticleName{Solitons in the Gauged Skyrme--Maxwell Model}

\ArticleName{Solitons in the Gauged Skyrme--Maxwell Model\footnote{This paper is a~contribution to the Special Issue on Topological Solitons as Particles. The~full collection is available at \href{https://www.emis.de/journals/SIGMA/topological-solitons.html}{https://www.emis.de/journals/SIGMA/topological-solitons.html}}}

\Author{Leandro Roza LIVRAMENTO~$^{\rm a}$, Eugen RADU~$^{\rm b}$ and Yakov SHNIR~$^{\rm ac}$}

\AuthorNameForHeading{L.R.~Livramento, E.~Radu and Y.~Shnir}

\Address{$^{\rm a)}$~BLTP, JINR, Dubna 141980, Moscow Region, Russia}
\EmailD{\href{mailto:livramento@theor.jinr.ru}{livramento@theor.jinr.ru}, \href{mailto:shnir@theor.jinr.ru}{shnir@theor.jinr.ru}}

\Address{$^{\rm b)}$~Department of Mathematics, University of Aveiro and CIDMA,\\
\hphantom{$^{\rm b)}$}~Campus de Santiago, 3810-183 Aveiro, Portugal}
\EmailD{\href{mailto:eugen.radu@ua.pt}{eugen.radu@ua.pt}}

\Address{$^{\rm c)}$~Institute of Physics, Carl von Ossietzky University Oldenburg, 26111 Oldenburg, Germany}

\ArticleDates{Received February 01, 2023, in final form June 03, 2023; Published online June 14, 2023}

\Abstract{We consider soliton solutions of the ${\rm U}(1)$ gauged Skyrme model with the pion mass term. The domain of existence of gauged Skyrmions is restricted from above by the value of the pion mass. Concentrating on the solutions of topological degree one, we find that coupling to the electromagnetic field breaks the symmetry of the configurations, the Skyrmions carrying both an electric charge and a magnetic flux, with an induced dipole magnetic moment. The Skyrmions also possess an angular momentum, which is quantized in the units of the electric charge. The mass of the gauged Skyrmions monotonically decreases with increase of the gauge coupling.}

\Keywords{Skyrme--Maxwell model; gauged Skyrmions; topological solitons; Skyrmions}

\Classification{81R12; 81T10}

\renewcommand{\thefootnote}{\arabic{footnote}}
\setcounter{footnote}{0}

The study of the classical soliton solutions in field theory can be
traced back to the pioneering paper by Skyrme \cite{skyrme,Skyrme:1962vh} (for a review, see \cite{Brown:2010api,Manton:2022,Manton:2004tk,Shnir:2018yzp}).
The Skyrme model was introduced in 1961 as a simple version of the nonlinear
sigma model in $3+1$ dimensions, which can be used as an effective theory of
atomic nuclei. It has been shown by Witten \cite{Witten:1983tw,Witten:1983tx} that the Skyrme model can be
derived from the $1/N_c$ expansion of the QCD low-energy effective Lagrangian.
The Skyrmions are topological solitons, in this framework the topological charge
 corresponding to the
physical baryon number.

 The simplest (and original) version
of the Skyrme model can be constructed for the ${\rm SU}(2)$ valued chiral field.
Then the model contains only three free parameters which
set the length and energy scales and the mass of the pion field, respectively. An appropriate fitting of these
parameters
 together with the
assumption that the slowly rotating Skyrmion can be considered as rigid body, allows to evaluate
various quantities like, e.g., mean square radii, $g$-factors of nucleons, and their magnetic
moments \cite{Adkins:1983ya}.
It turns out that the
agreement with the corresponding experimental data is surprisingly better
 than
what one would expect, being
within reasonable accuracy for the usual choice of the values of the parameters (for a review see \cite{Brown:2010api, Zahed:1986qz}).

However, the standard version of the Skyrme model has limited success, as
there are several problems with description of the nuclear
masses. First, in order to describe the properties of the pion excitations, the model must be supplemented with a
potential \cite{Adkins:1983hy,Battye:2004rw,Kopeliovich:2005vg}.
 Various modifications of the model's potential were considered,
which
typically do not much affect
the binding energy of the solitons, but may produce a
dramatic change of the shape of Skyrmions
\cite{Battye:2006na,Battye:2006tb,Dupuis:2018utr,Gudnason:2016mms,Gudnason:2015nxa,Gudnason:2016cdo,Gudnason:2016yix,Livramento:2022zly}.

The soliton solutions of the original Skyrme model
do not attain the topological bound, which yields a linear
relation between the static energy of the Skyrmions and their topological charges,
the interaction energy of the Skyrmions being
about 10$\%$ of the mass of a single Skyrmion. On the contrary,
the corresponding experimental data for nuclear binding energies are much lower, typically of the order of 1$\%$
of the mass of the nucleon. Since all binding energies vanish as the model approaches the topological bound,
which is often referred to as the BPS bound,
several modifications of the Skyrme model were proposed recently to improve the situation. The most radical version of the
BPS Skyrme theory is constructed via truncation of the model,
it contains only
a sixth-order derivative term, which is the topological
current density squared, and a potential \cite{Adam:2013wya, Adam:2010fg}.
It has been also suggested to consider a modification of
the Skyrme model
%to the form
which supports self-duality equations \cite{Ferreira:2021ryf, Ferreira:2017yzy,Ferreira:2017bsr}.
Another direction
is related with
extensions of the Skyrme model via inclusion of higher-order derivative terms
\cite{Adam:2013wya,Adam:2010fg,Adam:2013tga,Floratos:2001ih,Gudnason:2017opo,Marleau:1990nh,Marleau:1991jk},
with various modifications of the
potential and couplings.

On the other hand, addition of various terms simulating the contributions of scalar and vector mesons, also can lower the
binding energies of Skyrmions
\cite{Adkins:1983nw,Forkel:1991mv,Gudnason:2020arj,Jackson:1985yz,Naya:2018kyi,Naya:2018mpt,Schwesinger:1986xv,Yabu:1989un}.
In was pointed out by Sutcliffe \cite{Sutcliffe:2010et,Sutcliffe:2011ig,Sutcliffe:2015sta} that the BPS Skyrme
model coupled to an infinite
tower of vector mesons can be derived from the $4+1$ dimensional Yang--Mills
theory via the Atiyah--Manton construction \cite{Atiyah:1989dq}. This approach, inspired by the
holographic construction of the Skyrme model by Sakai and Sugimoto \cite{Sakai:2004cn},
provide a good approximation to the truncated BPS Skyrme theory.

Another extension of the Skyrme model can be obtained via ${\rm U}(1)$ gauging of the Skyrme field
\cite{Piette:1997ny,Radu:2005jp}.
This modification was originally motivated by construction of a semiclassical model of
Rubakov--Callan effect~\cite{Callan:1982ah, Rubakov:1981rg},
a process of monopole-catalyzed proton decay~\cite{Callan:1983nx,Kleihaus:1999ea}. It was pointed out that
the coupling to the electromagnetic field gives the Skyrmions the electric charge, the electrostatic repulsion decreases the total energy of the configuration although the binding energy of gauged Skyrmions in a minimal version of the Skyrme--Maxwell model increases~\cite{Piette:1997ny}.
Notably, the Skyrme--Maxwell theory also can be derived in a holographic model via an expansion of a Yang--Mills field of calorons~\cite{Cork:2021ylu}.

Let us remark that, however, the properties of the ${\rm U}(1)$ gauged Skyrmions have been better understood in lower
dimensions. As a $(2+1)$-dimensional analog of Skyrmions, the
so-called {\it ``baby Skyrmions"} were proposed in
a non-lineal ${\rm O}(3)$ sigma model with a fourth-order derivative term in $2+1$ dimensions
\cite{Bogolubskaya:1989ha,Bogolyubskaya:1989fz,Leese:1989gi}. This simplified theory emulates the
conventional Skyrme model in many respects. In particular,
it was used to study dynamics of the solitons \cite{Leese:1989gi, Piette:1994mh}
and the isorotations of multisoliton configurations without any restrictions of symmetry
\cite{Battye:2013tka,Halavanau:2013vsa}.
The planar Skyrme--Maxwell model was considered in \cite{Gladikowski:1995sc,Samoilenka:2015bsf,Shnir:2015twa}.
An interesting observation is that
in the strong coupling regime the magnetic flux coupled to the Skyrmion, is quantized, although there is no
topological reason of that.
 Furthermore, electrically charged planar Skyrmions were studied is an
extended model with a Chern--Simons term \cite{Navarro-Lerida:2018giv,Navarro-Lerida:2018siw, Samoilenka:2016wys}.
The gauged BPS baby-Skyrme model was considered recently in \cite{Adam:2017ahh,Casana:2019ztm}. Interesting examples of
gauged topological
solitons were also constructed in the ${\rm O}(3)$ sigma model \cite{Schroers:1995he},
and in the ${\rm U}(1)$ gauged Faddeev--Skyrme--Maxwell theory \cite{Samoilenka:2018oil, Shnir:2014mfa}.

 Returning to the (usual) Skyrme model in $3+1$ dimensions,
it was pointed out that a small contribution of the
energy of electromagnetic interaction in the Skyrme model may contribute to a tiny
difference between the masses of the neutron and the proton
($m_n = 939.566$~MeV, $m_p = 938.272$~MeV, respectively) \cite{Durgut:1985mu,Ebrahim:1987mu}.
Usually, the proton-neutron mass difference in the Skyrme model can be
explained via some mechanism of the spin-isospin symmetry breaking
associated with violation of the spherical symmetry of the soliton, with possible contribution of the electromagnetic interaction~\cite{Kaelbermann:1986ne}.
Another approach
is related with isospin symmetry breaking induced via coupling of the Skyrme field to vector mesons~\cite{Jain:1989kn,Schechter:1999hg}. This effect can be modeled by adding to the Lagrangian
an explicit symmetry breaking term~\cite{Speight:2018zgc} or via additional derivative terms~\cite{Sutcliffe:2023hdt}.
We argue below that the coupling to the electrostatic field alone will
break the symmetry of the Skyrme field, even in the sector of topological degree one.

In this paper we consider axially symmetric parametrization of the ${\rm U}(1)$ gauged Skyrmion of topological degree one. The existence of such solutions relies on the presence of the pion mass potential term, both the gauge coupling and electrostatic potential are restricted from above by the value of the pion mass.

Our results show that
the coupling to the electromagnetic field may deform the configurations. We shall construct explicit examples of electrically charged axially symmetric Skyrmions coupled to a circular magnetic flux and investigate their properties.

This paper is organised as follows. In Section \ref{sec:1}, we introduce the model, define both the topological and electric charges and consider linearized perturbations of the fields. In Section~\ref{sec:2}, we present the axially-symmetric ansatz for the Skyrme field and for the electromagnetic field. In Section~\ref{sec:3}, we define the boundary conditions to be used in finding the numerical solutions. In Section~\ref{sec:4}, we discuss the numerical results. Finally, in Section~\ref{sec:5}, we present our conclusions and further remarks.

\section{The model}\label{sec:1}

The basic and original version of the Skyrme model in $(3+1)$-Minkowski spacetime is defined
by the Lagrangian \cite{skyrme,Skyrme:1962vh}
\begin{gather}\label{Lag00}
{\mathcal L}_{{\rm Sk}}=\frac{f_\pi^2}{16} \tr \big(\pa_\mu U \pa^\mu U^\dagger\big)+\frac{1}{32 a_0^2} \tr
\big(\big[\pa_\mu U U^\dagger, \pa_\nu U U^\dagger\big]^2\big),
\end{gather}
where the Skyrme field $U(x)$ takes values in $S^3$, the group manifold of ${\rm SU}(2)$. In the context of application of the Skyrme
model as a candidate model of nuclear physics, the parameter $f_\pi$ may be interpreted as the
pion decay constant and $a_0$ is a dimensionless constant which can be defined from experimental data.
The simple model \eqref{Lag00} enjoys the chiral ${\rm SU}(2)_L\times {\rm SU}(2)_R$ global internal symmetry which acts of the Skyrme field
through the action $U \rightarrow g_L U g_R^{-1}$,
$\forall g_L, g_R \in {\rm SU}(2)$. In addition, the flat metric corresponds to $\eta_{\mu\nu}={\rm diag}(1,-1,-1,-1)$,
with
the Cartesian coordinates
$(x,y,z)$,
while $(r,\theta,\varphi)$
are the spherical ones.

The Lagrangian of the Skyrme model, in its minimal form \eqref{Lag00},
has just two terms, a usual sigma model term quadratic in derivatives and a Skyrme term quartic in derivatives.
A~potential term can be included in the Skyrme model, the most common choice
 being the pion mass term~\cite{Adkins:1983hy}
\begin{gather*}
V=\frac{m_\pi^2 f_\pi^2}{8} \tr (U-\one ) .
%\label{pot}
\end{gather*}
Here the parameter $m_\pi$ defines the asymptotic decay of the Skyrme field, its value
 being
calibrated to
fit the physical pion mass. The standard values of the parameters of model
are fixed to make contact with the experimental data for protons and pions~\cite{Adkins:1983hy}:
$f_\pi=108$~MeV, $a_0=4.84$ and $m_\pi=138$~MeV.
The inclusion of the potential term stabilizes the model with respect to isoratations,
beyond the rigid-body approximation~\cite{Battye:2014qva, Battye:2005nx}.

The requirement of finiteness of energy leads to restriction that the matrix-valued field $U$ approaches the vacuum
at all points at spatial infinity, $U \xrightarrow[\vec r \to\infty]{} {\one}$. This boundary condition compactifies
the domain space $\mathbb{R}^3 \to S^3$ and breaks the full chiral ${\rm SU}(2)_L\times {\rm SU}(2)_R$ symmetry to the diagonal subgroup.
The Skyrme field becomes a map $U\colon S^3 \mapsto S^3$ and the corresponding topological degree $B$ can be written as
\begin{gather}
B=-\frac{1}{24 \pi^2} \int {\rm d}^3x \varepsilon_{ijk} \tr (R_i R_j R_k) . \label{chargemain}
\end{gather}
where $R_i=(\partial_i U)U^\dagger$ is the $\mathfrak{su}(2)$-valued current. The charge~\eqref{chargemain} is interpreted as the
baryon number.

One can construct a gauged version of the Skyrme model by gauging of the ${\rm U}(1)$ subgroup of the ${\rm SU}(2)$ global
symmetry, associated with the generator of its Cartan subalgebra, see~\cite{Callan:1983nx,Piette:1997ny,Radu:2005jp}.
The covariant derivative of the Skyrme field is defined as
\[
D_\mu U = \pa_\mu U -{\rm i} e {\mathcal A}_\mu [Q, U ],
\]
where the charge matrix is $Q\equiv\frac{1}{2}\big(\frac{1}{3} \mathbb{I}+\tau_3\big)={\rm diag} \big(\frac{2}{3}, -\frac{1}{3}\big)$.
Then the full Lagrangian of the ${\rm U}(1)$ gauged Skyrme model can be written as
\begin{gather}
{\mathcal L}=\frac{f_\pi^2}{16} \tr \big(D_\mu U D^\mu U^\dagger\big)+
\frac{1}{32 a_0^2} \tr \big(\big[D_\mu U U^\dagger, D_\nu U U^\dagger\big]^2\big)\nonumber\\
\hphantom{{\mathcal L}=}{}+
\frac{m_\pi^2 f_\pi^2}{8} \tr (U-\one ) -\frac{1}{4} {{\mathcal F}}_{\mu\nu} {{\mathcal F}}^{\mu\nu},\label{Lag0}
\end{gather}
while the electric charge is $Q_e = \int {\rm d}^3x \pa_i {\mathcal F}^{i0}$.
It is convenient to scale away the parameters~$f_\pi$,~$a_0$ by introducing
the energy and length scales $f_\pi/(4 a_0)$ and $2/(a_0 f_\pi)$, respectively. The rescaled pion mass parameter is
$m=2 m_\pi/(a_0 f_\pi)$ and we absorb the parameter $a_0$ into the redefined gauge potential,
$A_\mu \equiv a_0 {\mathcal A}_\mu$, $F_{\mu\nu} \equiv a_0 {\mathcal F}_{\mu\nu} $ with the gauge coupling $g\equiv e/a_0$.

In terms of these units the Skyrme--Maxwell Lagrangian \eqref{Lag0} becomes
 \begin{gather}\label{Lag0b}
{\mathcal L}=\frac{1}{2} \tr \big(D_\mu U D^\mu U^\dagger\big)+\frac{1}{16} \tr
\big(\big[D_\mu U U^\dagger, D_\nu U U^\dagger\big]^2\big)+m^2 \tr(U-\one)
-\frac{1}{2} {{F}}_{\mu\nu} {{F}}^{\mu\nu} .
\end{gather}
In the static gauge the electric charge is now reduced to
\begin{gather}
Q_e = -\frac{1}{a_0}\int {\rm d}^3x \pa_i^2 A_0=-\frac{1}{a_0}\oint {\rm d}\vec{S}\cdot \vec{\nabla} A_0. \label{Qelectric}
\end{gather}

\subsection{Linearized perturbations}

The Skyrme field can be decomposed into the scalar meson field $\phi_0$ and the pion isotriplet
$\phi_k$ via
\begin{gather}
U=\phi_0 \one+{\rm i} \sum_{k=1}^3\phi_k \tau_k, \label{decomposition}
\end{gather}
where $\tau_k$ denotes the triplet of usual Pauli matrices, and the field $\phi^a = (\phi_0,\phi_k)$ is restricted
to the unit sphere, $\phi^a \cdot \phi^a = 1 $.
In these component notations the Lagrangian for the ${\rm U}(1)$ gauged Skyrme model \eqref{Lag0b} can then be written as
\begin{gather}
{\mathcal L}=- \frac{1}{2} F_{\mu\nu} F^{\mu\nu}+ D_\mu \phi^a D^\mu \phi^a
-\frac12 \big(D_\mu \phi^a D^\mu \phi^a\big)^2 +
\frac{1}{2} \big(D_\mu \phi^a D_\nu \phi^a\big)\big(D^\mu \phi^b D^\nu \phi^b\big)\nonumber\\
\hphantom{{\mathcal L}=}{} - 2 m^2 (1-\phi_0) ,\label{Lag}
\end{gather}
where
\begin{gather*}
D_\mu \phi^\alpha = \pa_\mu \phi^\alpha -g A_\mu \varepsilon_{\alpha\beta} \phi^\beta ,
\qquad D_\mu \phi^A = \pa_\mu \phi^A, \qquad \alpha, \beta=1, 2,\quad A=0, 3.
%\label{covariant}
\end{gather*}
In other words, the gauged Skyrme model \eqref{Lag} is invariant with respect to the local ${\rm U}(1)$ gauge transformations{\samepage
\begin{gather}
U\rightarrow {\rm e}^{{\rm i} g \frac{\alpha}{2}\tau_3} U {\rm e}^{-{\rm i} g \frac{\alpha}{2}\tau_3} , \qquad {\rm or}\qquad
\phi_1+{\rm i} \phi_2\rightarrow {\rm e}^{-{\rm i} g \alpha} (\phi_1+{\rm i} \phi_2),
\qquad A_\mu \rightarrow A_\mu + \pa_\mu \alpha,
\label{gauge}
\end{gather}
where $\alpha$ is any real function of coordinates.}

The vacuum of \eqref{Lag} corresponds to $U=\one$, $D_\mu \phi^a =0$ and $F_{\mu\nu}=0$. In the static gauge one can consider the vacuum boundary conditions
\begin{gather}
U(\infty)=\one,\qquad A_0(\infty) = V,\qquad A_i(\infty) = 0 ,\label{asympcondition}
\end{gather}
where $V$ is a real constant. Note that the asymptotic value of the electric potential $A_0(\infty)$ can be adjusted via
residual ${\rm U}(1)$ degrees of freedom, in particular, the transformations~\eqref{gauge} with $\alpha = -V t$ allows
us to set $A_0(\infty)=0$. In such a gauge the components of the
charged pions field transform as $\phi_\alpha \rightarrow {\rm e}^{{\rm i} \omega t} \phi_\alpha$, with $\omega=g V$.
In other words, in the Skyrme--Maxwell model \eqref{Lag0} the isorotations of the Skyrmions are associated with
the time-dependent gauge transformations~\cite{Radu:2005jp}. Hereafter we will fix the boundary conditions~\eqref{asympcondition}
 setting $\omega =0$.

The asymptotic expansion of the fields around the vacuum \eqref{asympcondition} yields
$U_{{\rm ex}} \sim (1-v_0) \one+{\rm i} v_k \tau_k$ and ${A}_\mu=a_\mu+V \delta_{0\mu}$,
where $v_k$ and $a_\mu$ are perturbative excitations of the triplet of pions fields and the electromagnetic potential, respectively,
while $v_0$ is the field excitation of the scalar component~$\phi_0$. Since about the vacuum
$\phi_0\sim(1-v_0)$, $\phi_k \sim v_k$, the constraint on the
components of the scalar field yields $v_k^2=1-(1-v_0)^2$. Thus, $v_k^2 \approx 2 v_0+ O\big(v_0^2\big)$ and $\pa_i v_0 \approx v_k
\pa_i v_k +O\big(v_0^2, v_0 \pa_i v_0\big)$ and the fluctuations of the $v_0$-field correspond to the second order of
expansion of the fields around the vacuum.

Hence, the linearized perturbations of the fields around the vacuum are
\begin{gather*}
U_{{\rm ex}} = \one + {\rm i} v_k \tau_k + O \big(v_k^2\big) ,\qquad A_\mu=a_\mu +V \delta_{0\mu} .
\label{flutuations}
\end{gather*}
Note that the perturbations of the pion mass term correspond to the second order of expansion of the
fields $v_k$,
since
$2 (1-\phi_0)\sim 2 v_0\sim v_k^2+O\big(v_k^3\big)$,
and the linearized
perturbation of the term
$D_\mu \phi^a$
yields a leading contribution to the dynamics of the scalar excitations.
Then the asymptotic expansion of the Lagrangian~\eqref{Lag} in the static gauge gives
\begin{gather*}%\label{Lagex}
{\mathcal L}_{{\rm pert }}=-\big[f^2 + (\pa_i v_k)^2 +m^2 v_k^2 -
g^2 V^2 \big(v_1^2+v_2^2\big)\big],
\end{gather*}
where $f^2=(\pa_i a_0)^2 -\pa_i a_j (\pa_i a_j-\pa_j a_i) $ is the contribution of the Maxwell term.
The corresponding linearized equation for the fluctuations of the scalar fields~$v_a$ is
\begin{gather*}%\label{Lagexn}
\pa_i^2 v_a -\big[m^2 v_a -g^2 V^2 (v_1 \delta_{a1}+v_2 \delta_{a2})\big]=0.
\end{gather*}
Therefore, the effective mass of charged pions, associated with the excitations
$v_\pm = \frac{1}{\sqrt 2}(v_1 \pm v_2)$, is $m_{\rm eff}^\pm=\sqrt{m^2-g^2 V^2}$, while the mass of the
uncharged component, associated with excitation of the $\phi_3$ field, is slightly higher,\footnote{This is not in especially
good agreement with the experimental values of physical masses of pions, $m_{\pi_\pm}=139.570$~MeV, $m_{\pi_0}=134.977$~MeV. However,
we can expect that the quantum corrections to the masses of excitations,
which will also take into account a quartic pion interaction and contributions from
vector mesons, may improve the situation.} $m_{\rm eff}^{(v_3)}=m$.

Localized massive scalar modes with exponentially decaying tail may exist if $m>0$ and
\begin{gather*}
|g V|\leq m .%\label{bound}
\end{gather*}
In the critical case $|g V| =m$ the asymptotic of charged modes posses a dipole as a leading contribution, similar to the
neutral mode $v_3$ in the massless limit. Thus, the pattern of interaction between gauged Skyrmions becomes rather involved, it
may include both long-range and dipole forces, as well as short-range Yukawa interactions.

\subsection{Topological density and the stress-energy tensor}

The ${\rm U}(1)$ gauge covariant generalization
of the usual topological charge \eqref{chargemain} associated to the transformation \eqref{gauge}
can be constructed by replacing $\pa_i \rightarrow D_i$ in the expression \eqref{chargemain},
i.e.,
\begin{gather*}
B_g=-\frac{1}{24 \pi^2} \int {\rm d}^3x \varepsilon_{ijk} \tr \big(D_i U U^{-1} D_j U U^{-1} D_k U U^{-1}\big).
%\label{comutden}
\end{gather*}
and then subtracting the functional $B_{\rm mag}=\int {\rm d}^3 x\frac{{\rm i} g}{32 \pi^2} (\varepsilon_{ijk} {F}_{jk}) \tr \big( {\tau_3, \pa_i U } U^{-1}\big)$. Due the fact that $B_g$ and $B_{\rm mag}$ are invariant by the transformations~\eqref{gauge}, the same follows for the topological charge of the Skyrme--Maxwell model, which is so defined by
\[
Q=B_g-B_{\rm mag} = B + \int {\rm d}^3x \pa_i \Lambda_i ,
%\label{densityrelpre}
\]
where
\[
\Lambda_i =
-\frac{{\rm i} g}{16 \pi^2} \varepsilon_{ijk} {A}_j \tr \big( \{\tau_3, \pa_k U \} U^{-1}\big).
\]
Thus, the topological charge of the Skyrme--Maxwell model is
an usual winding number~\eqref{chargemain}, plus a surface term which only depends on the boundary conditions of the fields \cite{Cork:2021ylu, Piette:1997ny}. In the abelian Skyrme--Maxwell model~\eqref{Lag0} with the boundary conditions \eqref{asympcondition} the flux of $\Lambda_i$ is vanishing\footnote{However, in the ${\rm SU}(2)$ gauged Skyrme model, it can unwind the Skyrmion \cite{Brihaye:2002nz, Brihaye:2001qt}. This yields a simple classical model of monopole catalysis
of nucleon decay (the Rubakov--Callan effect).} \cite{Piette:1997ny}.

Therefore, the topological charge of the ${\rm U}(1)$ gauged Skyrmions is still defined by the
Skyrme map \eqref{chargemain}, i.e.,
\[
Q= B.%\label{densityrel}
\]

In terms of the decomposition~\eqref{decomposition}, the topological charge is expressed as
\begin{gather}
B= \int {\rm d}^3x q(\vec r)= -\frac{1}{12 \pi^2} \int {\rm d}^3x \varepsilon_{abcd} \varepsilon_{ijk} \phi^a \pa_i \phi^b \pa_j \phi^c \pa_k \phi^d ,\nonumber\\
\Lambda_i =-\frac{g}{4 \pi^2} \varepsilon_{ijk} {A}_j \varepsilon_{AB} \phi_A \pa_k\phi_B ,\label{lam}
\end{gather}
where $q(\vec r)$ is the topological charge density and the indices $A, B=0, 3$.

The stress-energy tensor of the model
\eqref{Lag0} can be obtained by variation of the action with respect to Minkowski metric
$\eta^{\mu \nu}$, it gives
\[
T^{\mu\nu}=T^{\mu\nu}_{\rm (M)}+T^{\mu\nu}_{\rm (S)},%\label{T}
\]
where the electromagnetic contribution of the Maxwell term is
\[
T^{\mu\nu}_{\rm (M)}=-2 F^{\mu\sigma } F^{\nu}_{\:\:\sigma}+\frac{\eta^{\mu\nu}}{2} F_{\alpha\beta} F^{\alpha\beta},
\]
and the stress-energy tensor of the U(1)-gauged Skyrmions is
\begin{gather}
\nonumber T^{\mu\nu}_{\rm (S)} = 2 \big[ D^\mu \phi_a D^{\nu}\phi^a-\big(D^{[\mu} \phi^a D^{\alpha]} \phi^b\big) \big(D^{[\nu} \phi_a D_{\alpha]}\phi_b\big)\big] \\
\hphantom{T^{\mu\nu}_{\rm (S)} =}{}
-\eta^{\mu\nu} \left( ( D_\alpha \phi_a )^2-\frac12 ( D_{[\alpha} \phi_a D_{\beta]} \phi_b )^2- 2 m^2 (1-\phi_0)\right) \label{stress}.
\end{gather}

In the static gauge the Hamiltonian of the Skyrme--Maxwell model can be
written as
\begin{gather}
{\mathcal H}_{{\rm static}}={\mathcal H}_{1} + {\mathcal H}_{2} ,\label{staticL}
\end{gather}
where ${\mathcal H}_1$ and ${\mathcal H}_2$ are non-negative terms given by
\begin{gather*}
{\mathcal H}_{1}= \frac{1}{2} |F_{ij}|^2+|D_i \phi_a|^2
+\frac12 |D_{[i} \phi_a D_{j]} \phi_b |^2+ 2 m^2 (1-\phi_0) ,\\
{\mathcal H}_2=|\pa_i A_0|^2+g^2 A_0^2 M_\phi^2,\qquad {\rm with}\quad M_\phi^2\equiv \left[\big(1+|\pa_i \phi_A|^2\big) |\phi_\alpha|^2+\frac{1}{4} \big|\pa_i \big(|\phi_A|^2\big)\big|^2\right] .%\label{H12}
\end{gather*}
where $|\pa_iA_0|^2=\pa_iA_0 \pa_iA_0$, and so on. Clearly, the last term in ${\mathcal H}_2$ yields the Gauss law. The function $g^2 M_\phi^2$ behaves like a spatially dependent square mass for the electric potential, which appears in the interior of the Skyrmion and vanishes asymptotically. The remaining gauge degrees of freedom can be fixed imposing the Coulomb gauge, $\pa_i A_i=0$. Then the static Maxwell equations for the electric and magnetic potentials can be written as
\[
\pa_j^2 A_0 = g^2 M_\phi^2 A_0 ,\qquad \pa_j^2 A_i = -g \varepsilon_{\alpha\beta} \phi_\beta \big[(1+ D_j \phi_a D_j \phi_a ) D_i \phi_\alpha - (D_j \phi_a D_j \phi_\alpha) D_i \phi_a
\big] .%\label{A0eq}
\]

\section{Axially symmetric ansatz and effective Lagrangian}\label{sec:2}

An obvious
correspondence between the gauge transformations \eqref{gauge} and isorotations of the components of the Skyrme field
suggests that, in general, gauged Skyrmion of topological degree one is not spherically symmetric \cite{Piette:1997ny}. Hence we
consider a general axially symmetric ansatz
(see \cite{Battye:2005nx,Ioannidou:2006nn,Krusch:2004uf,Radu:2005jp}):
\begin{gather}\label{U}
 \phi_1+i \phi_2=\psi_1(r,\theta) {\rm e}^{{\rm i} n \varphi}, \qquad
 \phi_3=\psi_2(r,\theta) , \qquad \phi_0=\psi_3(r,\theta) ,
\end{gather}
where $n$ an integer (so-called ``vorticity'') and the sigma-model constraint $\psi_1^2+\psi_2^2+\psi_3^2=1$ is imposed. The gauge field is parameterized by the two potentials (magnetic and electric, respectively)
\begin{gather}\label{A}
A \equiv A_\mu {\rm d}x^\mu=A_\varphi (r,\theta) {\rm d}\varphi+A_0(r,\theta) {\rm d}t.
\end{gather}
All five functions which parameterize the ansatz \eqref{U}, \eqref{A} depend on the radial variable $r$ and the polar angle $\theta$.

Substituting the axial ansatz \eqref{U}, \eqref{A} into
the Hamiltonian of the Skyrme--Maxwell model \eqref{staticL} and into
the Lagrangian \eqref{Lag}, we obtain
\begin{gather}
L= -2 \pi \int {\rm d}\theta {\rm d}r r^2 \sin \theta\big[
{\mathcal F}^-+{\mathcal L}_{2}^{-}+{\mathcal L}_{4}^{-}+2 m^2 (1-\psi_3)\big] ,\label{lageff}\\
H= 2 \pi \int {\rm d}\theta {\rm d}r r^2 \sin \theta\big[{\mathcal F}^++{\mathcal L}_{2}^{+}+{\mathcal L}_{4}^{+}+2 m^2 (1-\psi_3)\big] ,\label{Heff}
\end{gather}
where
\begin{gather*}
{\mathcal F}^\mp \equiv \frac{1}{r^2 \sin^2 \theta}
\left(
A_{\varphi,r}^2+\frac{A_{\varphi,\theta}^2}{r^2}\right)
\mp
\left(A_{0,r}^2+\frac{A_{0,\theta}^2}{r^2}\right),
\\
{\mathcal L}_{2}^\mp \equiv
 \psi_{a,r}^2+\frac{\psi_{a,\theta}^2}{r^2}
+\psi_1^2
\left(
\frac{(n+g A_\varphi)^2}{r^2\sin^2 \theta}
\mp g^2 A_0^2
\right),
\\
{\mathcal L}_{4}^\mp \equiv \frac{1}{r^2}
\left[
(\psi_{3,\theta} \psi_{2,r} -\psi_{2,\theta} \psi_{3,r} )^2
+(\psi_{2,\theta} \psi_{1,r} -\psi_{1,\theta} \psi_{2,r} )^2
+(\psi_{3,\theta} \psi_{1,r} -\psi_{1,\theta} \psi_{3,r} )^2
\right]\\
\nonumber
\hphantom{{\mathcal L}_{4}^\mp \equiv}{}
+ \psi_1^2
\left(
\frac{(n+g A_\varphi)^2}{r^2 \sin^2 \theta}
\mp g^2 A_0^2
\right)
\left(
\psi_{a,r}^2+\frac{\psi_{a,\theta}^2}{r^2}\right) .
\end{gather*}
Here a comma denotes partial differentiation, i.e., $A_{\varphi,r}\equiv \frac{\partial A_\varphi}{\partial r}$, etc.
The corresponding
static field equations can be obtained from the variation of the effective Lagrangian \eqref{lageff} with respect to the functions parametrizing the ansatz~\eqref{U},~\eqref{A}. In particular,
one finds the
reduced Maxwell equations
for the magnetic and electric potentials:
\begin{gather} \pa_r^2 A_\varphi+\frac{\pa_\theta^2 A_\varphi-\pa_\theta A_\varphi \cot\theta}{r^2}=g (n+g A_\varphi) K ,
\label{eqAphi}
\\
\pa_i^2 A_0=g^2 A_0 K,\qquad {\rm with}\quad K \equiv \psi_1^2 \left[1+\psi_{a,r}^2+\frac{\psi_{a,\theta}^2}{r^2}\right] .\label{eqA0n}
\end{gather}
Further, the linearized equations for the pion fields for large $r$
become
\begin{gather*}%\label{Lagexnb}
\pa_r^2 \psi_1 + \frac{2 \pa_r \psi_1}{r} -\big(m^2-g^2 V^2\big) \psi_1=0,\qquad \pa_r^2 \psi_2 + \frac{2 \pa_r \psi_2}{r} - m^2 \psi_2=0,
\end{gather*}
where $V = A_0(r\to \infty)$.
Thus, the fields asymptotically decay as $\psi_1 \sim {\rm e}^{-\sqrt{m^2-g^2V^2}r}$ and $\psi_2 \sim {\rm e}^{-m r}$
and localized solutions exist if $|m| \ge |gV|$. Finally, substituting the ansatz \eqref{U} into the Skyrme topological charge \eqref{chargemain} gives
$B=n$ \cite{Shnir:2009ct}. Hereafter we restrict our consideration to the gauged Skyrmions of degree one setting $n=1$.

 The total mass-energy of the configuration is defined
as the volume integral over all space of the $T_{00}$
component of the energy-momentum tensor,
 $E=\int {\rm d}^3 x T_{0}^0 $,
 which also includes a purely electromagnetic contribution.
The total angular momentum of the gauged Skyrmion
is defined as
$J=\int {\rm d}^3 x T_{0}^{\varphi}$.
Using the field equations and the definition of the topological charge~\eqref{chargemain}, one can show that the angular momentum of the gauged Skyrmion is classically quantized in units set by the electric charge~\cite{Radu:2005jp}.
 Indeed, using the axially symmetric ansatz \eqref{U} and \eqref{A} the component $T_{0\varphi}$ of the stress-energy tensor~\eqref{stress} can be written as
\[
T_{0\varphi}= 2\big[\pa_i A_0 \pa_i A_\varphi+g A_0 (n+g A_\varphi) K\big],
\]
where $K$ is defined by \eqref{eqA0n}. However, using the Euler-Lagrange equation for the electric potential \eqref{eqA0n} we can write $T_{0\varphi}$ as a total derivative, i.e.,
\[
T_{0\varphi}= 2\left[\pa_i A_0 \pa_i A_\varphi+\left(\frac{n}{g}+A_\varphi\right) \pa_i^2 A_0\right]=2 \pa_i\left[\left(\frac{n}{g} +A_\varphi\right)\pa_i A_0\right] .
\]
Therefore, making use of the equation~\eqref{Qelectric}, the expression for the angular momentum of the gauged Skyrmion can be written as
\[
J= \int {\rm d}^3x T_{0}^{\varphi} =
- 2 \oint_\infty {\rm d}\vec{S}\cdot \vec{\nabla} A_0 \left(\frac{n}{g}+A_\varphi\right)=-\frac{2 n}{g} \oint_\infty {\rm d}\vec{S}\cdot \vec{\nabla} A_0 = 2 n a_0 \frac{Q_e}{g},
\]
where we take into account that
$\frac{n}{g}+A_\varphi \rightarrow \frac{n}{g}$, as $r\rightarrow \infty$. Since the dimensions of the energy and the length are given by $f_\pi/(4 a_0)$ and $2/(f_\pi a_0)$, respectively, the dimension of angular momentum is $1/\big(2 a_0^2\big)$. Using the definition $g=e/a_0$, we can write the angular momentum in the original units of the ${\rm U}(1)$ gauged Skyrme model \eqref{Lag0} as
$J=n \frac{Q_e}{e}$.

In addition to $M$, $J$
and $Q_e$, another quantity of interest
is the magnetic dipole moment~$\mu_{m}$,
which is computed from the far field expression of the magnetic potential,
$A_\varphi \to \frac{\mu_m}{r}\sin^2 \theta+O\big(r^{-2}\big)$.

Also,
as a measure of the deformation degree of the configurations, we consider the ratio
\begin{gather}
\epsilon = \frac{\sqrt{\langle x^2\rangle}}{\sqrt{\langle z^2\rangle}} , \qquad
{\rm with} \quad \langle x^2\rangle = \frac{\int {\rm d}^3 x x^2  q(\vec r)}{\int {\rm d}^3x q(\vec r)},\qquad
\langle z^2\rangle = \frac{\int {\rm d}^3 x z^2 q(\vec r)}{\int {\rm d}^3x q (\vec r)} ,
\label{deform}
\end{gather}
where ${\langle |\rho|\rangle}$ and
${\langle |z|\rangle}$ are the mean dimensions of the axially symmetric
gauged Skyrmion,
defined as averages over the topological charge density $q(\vec r)$ \eqref{lam}.

\section{Numerical scheme and the boundary conditions}\label{sec:3}

To find numerical solutions of field equations
we used the software package CADSOL based on the Newton--Raphson algorithm~\cite{schoen}.
The calculations are performed on an equidistant grid
in spherical coordinates $r$ and $\theta$. Typical grids we used have sizes $70 \times 60$.
In our numerical scheme we map the infinite interval of the variable $r$ onto the compact radial coordinate
$x=\frac{r}{1+r} \in [0:1]$. Estimated numerical errors are of order of $10^{-5}$.

The system of the field equations of the Skyrme--Maxwell model represents a set of five coupled elliptic partial differential
equations with mixed derivatives, to be solved numerically
subject to the appropriate boundary conditions. As usual,
they follow from the condition of regularity of the fields on
the symmetry axis and symmetry requirements, as well as
the condition of finiteness of the energy of the system.
Explicitly, we impose
\begin{gather*}
\psi_1 \big|_{r=0}=0,\qquad
\psi_2 \big|_{r=0}=0,\qquad
\psi_3 \big|_{r=0}=-1,\qquad
A_\varphi \big|_{r=0}=0,\qquad
\partial_r A_0 \big|_{r=0}=0,
\\
\psi_1 \big|_{r= \infty}=0,\qquad
\psi_2 \big|_{r= \infty}=0,\qquad
\psi_3 \big|_{r= \infty}=1,\qquad
A_\varphi \big|_{r= \infty}=0,\qquad
A_0 \big|_{r= \infty}=V,
\end{gather*}
and
\begin{gather*}
\psi_1 \big|_{\theta=0,\pi}=0,\qquad\!\!
\partial_\theta \psi_2 \big|_{\theta=0,\pi}=0,\qquad\!\!
\partial_\theta \psi_3 \big|_{\theta=0,\pi}=0,\qquad\!\!
A_\varphi \big|_{\theta=0,\pi}=0,\qquad\!\!
\partial_\theta A_0 \big|_{\theta=0,\pi}=0.
\end{gather*}
The solutions also possess a
reflection symmetry with respect to the equatorial plane, i.e.,
\begin{gather*}
\partial_\theta \psi_1 \big|_{\theta=\pi/2}\!=0,\qquad\!\!\!
\psi_2 \big|_{\theta=\pi/2}\!=0,\qquad\!\!\!
\partial_\theta \psi_3 \big|_{\theta=\pi/2}\!=0,\qquad\!\!\!
\partial_\theta A_\varphi \big|_{\theta=\pi/2}\!=0,\qquad\!\!\!
\partial_\theta A_0 \big|_{\theta=\pi/2}\!=0.
\end{gather*}

\section{Numerical results}\label{sec:4}

For a fixed value of the pion mass $m=1$
the solutions of the gauged Skyrme model
depend on two continuous parameters, the values the gauge coupling $g$
and the electric
potential at infinity $A_0(\infty)=V$.
Localized solutions of the model~\eqref{Lag}
exist as $|gV| \le m =1$.

In our approach,
we have first considered a computation of the ungauged $B=1$ spherically symmetric Skyrmion (with $g=V=0$), which
is used as an input for following numerical calculations.
Then both~$g$ and~$V$ are increased in small steps.
 Setting $V=0$ and increasing the gauge coupling constant~$g$
 yields a branch of gauged
 static
 Skyrmions coupled to a toroidal magnetic flux
 and $A_0\equiv 0$.
Notably, there is no purely electrically charged solutions, all configurations being equipped with a toroidal magnetic flux which induces a dipole magnetic moment of the gauged Skyrmion.

\begin{figure}[t]\centering
\includegraphics{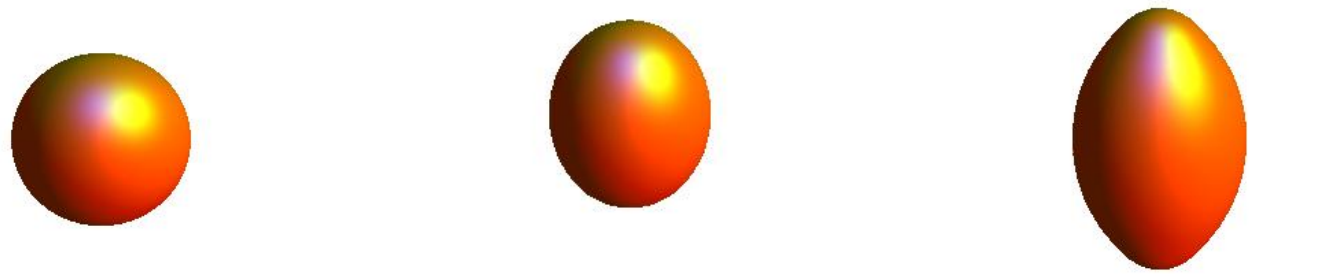}

\vspace{2mm}

\includegraphics{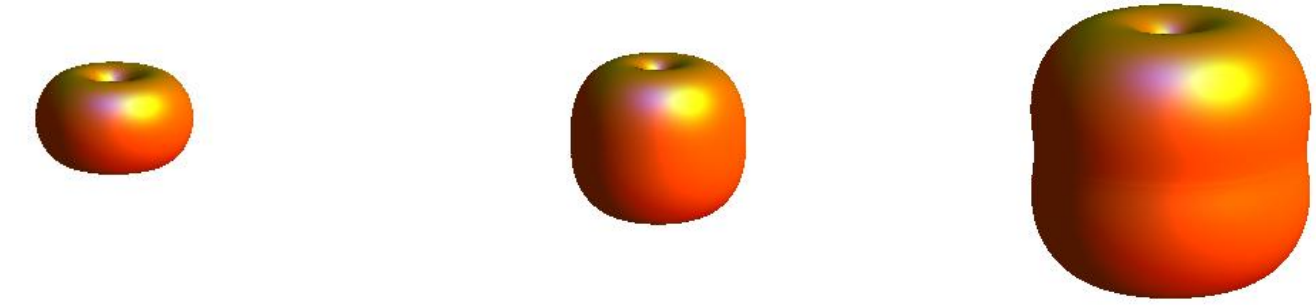}

\caption{Isosurfaces of the total energy density
distributions (top row) and the angular momentum density (bottom row) of the gauged $B=1$ Skyrmion for $g=0.1$
(left figures), $g=2$ (middle figures) and $g=10$ (right figures) at $V=0.1$ and $m=1$.}\label{eng-isosurfaces}
\end{figure}

\begin{figure}[h]\centering
\includegraphics[height=.33\textheight, angle =-90]{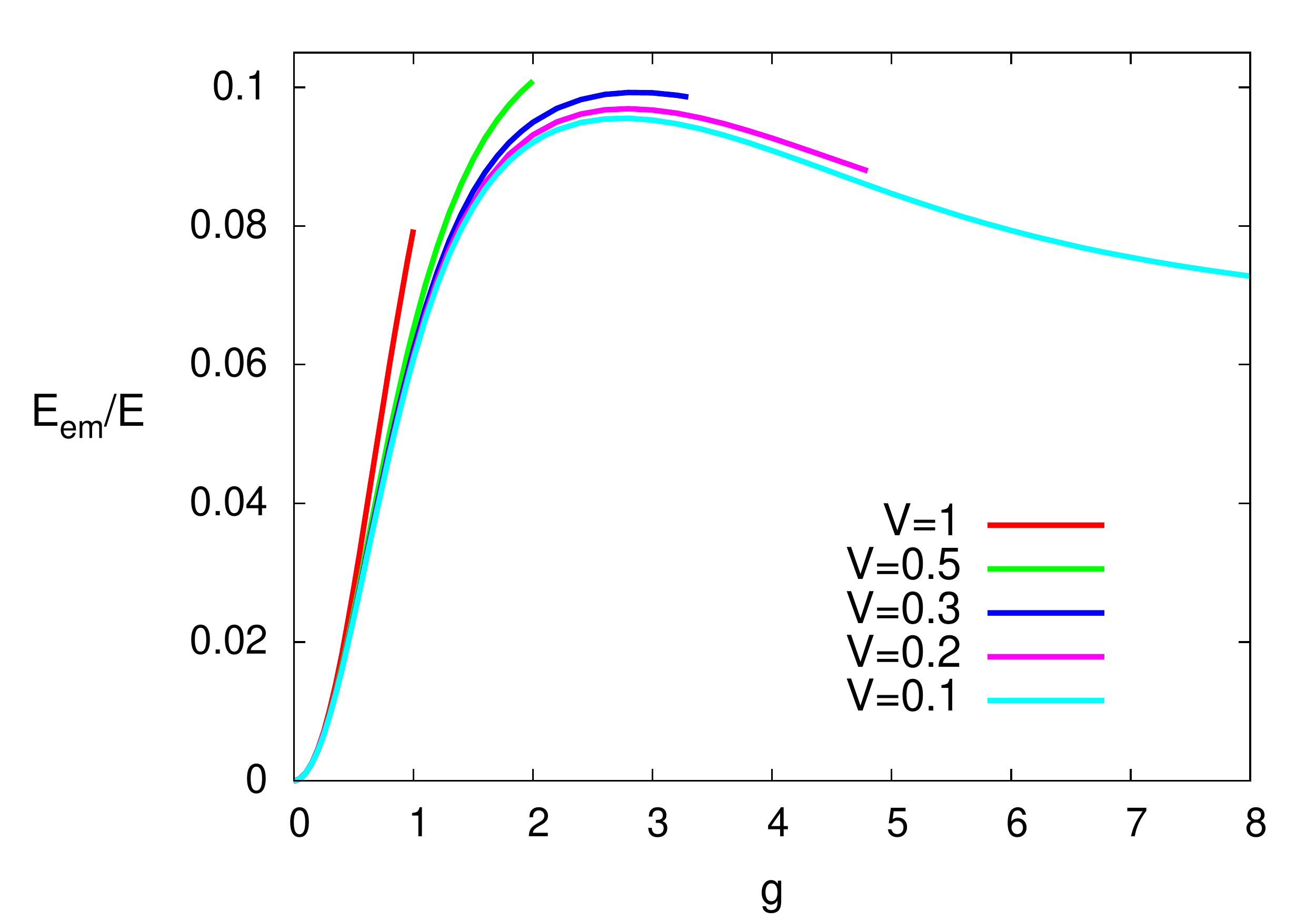}
\includegraphics[height=.33\textheight, angle =-90]{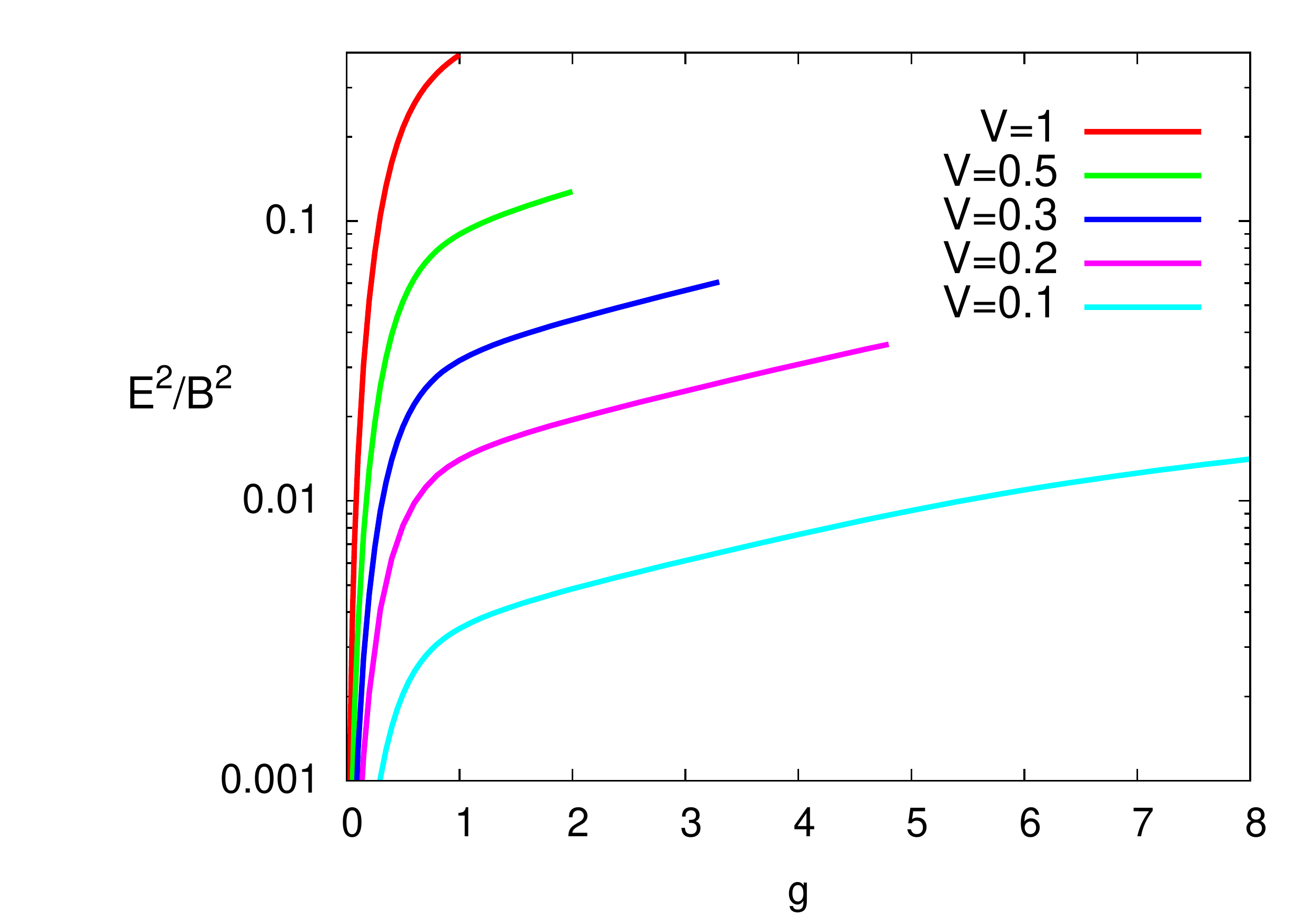}
\includegraphics[height=.33\textheight, angle =-90]{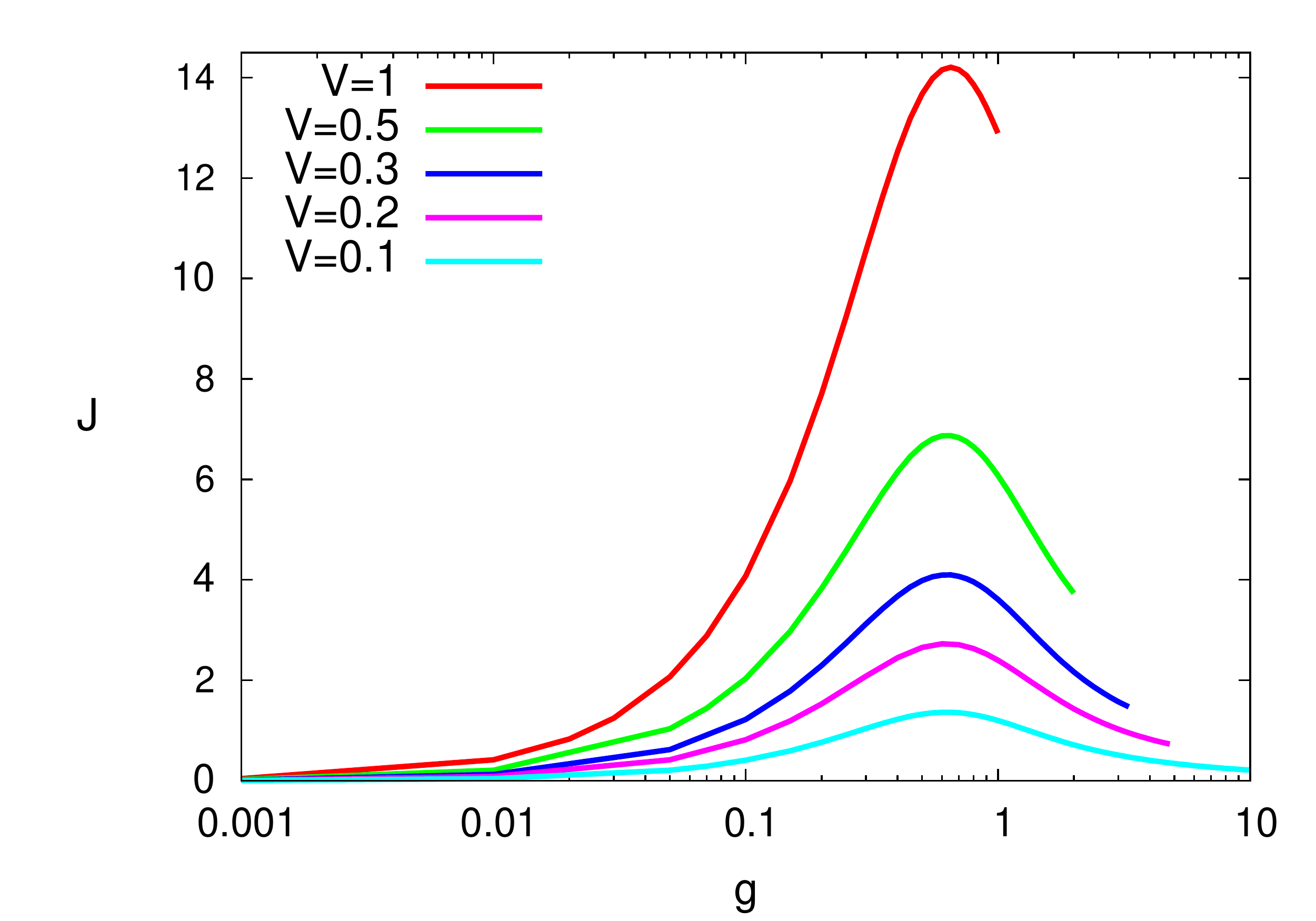}
\includegraphics[height=.33\textheight, angle =-90]{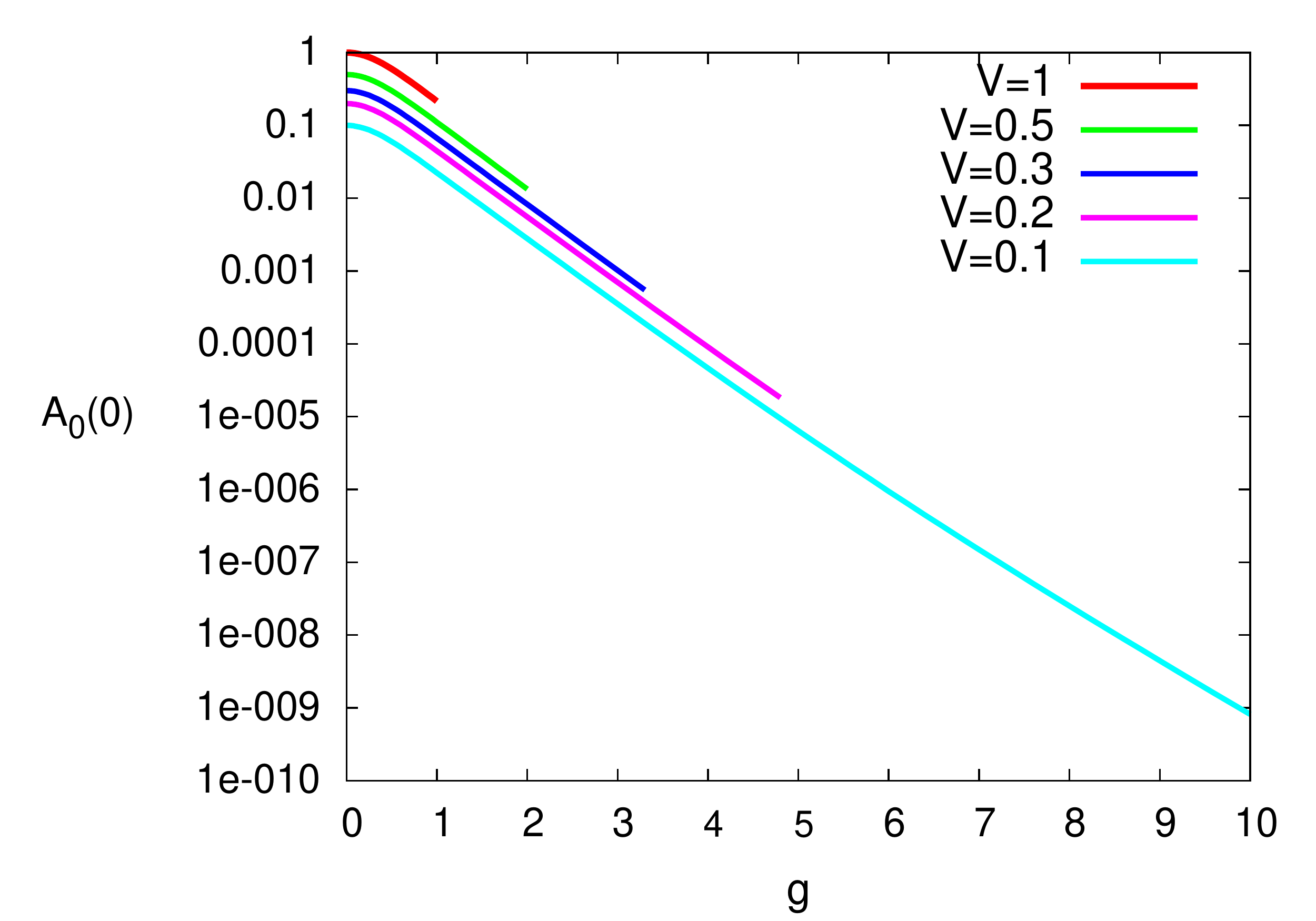}
\caption{Gauged $B=1$ Skyrmion: The ratio of the electromagnetic and total energies (upper left),
the ratio of the electric and magnetic energies (upper right), the total angular momentum (lower left) and
the value of the electric potential $A_0$ at the center of the Skyrmion (lower right) are plotted as
functions of the gauge coupling constant $g$ for some set of values of the parameter $V$ at $m=1$.}\label{eng-g}
\end{figure}

First, we confirm an observation \cite{Piette:1997ny} that for any non-zero value of the gauge coupling~$g$ the spherical symmetry of the $B=1$ ungauged Skyrmion is broken, the ${\rm U}(1)$ gauged Skyrmions are axially symmetric. In Figure~\ref{eng-isosurfaces}, we exhibited some typical examples of
the isosurfaces of the energy density and the angular momentum density at $V=0.1$ and $m=1$ and some set of values of the gauge coupling.
\begin{figure}[h]\centering
\includegraphics[height=.215\textheight]{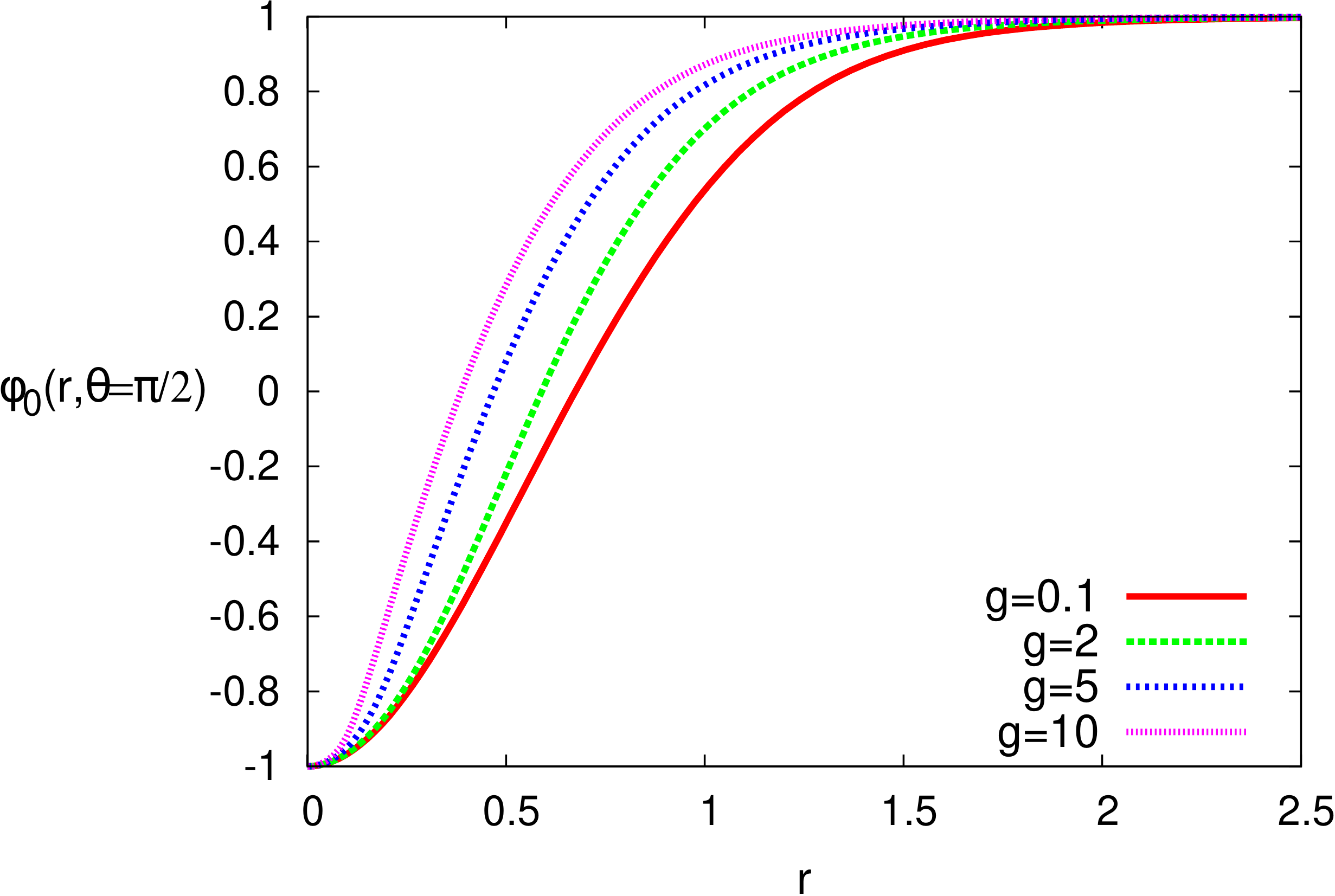}\qquad
\includegraphics[height=.215\textheight]{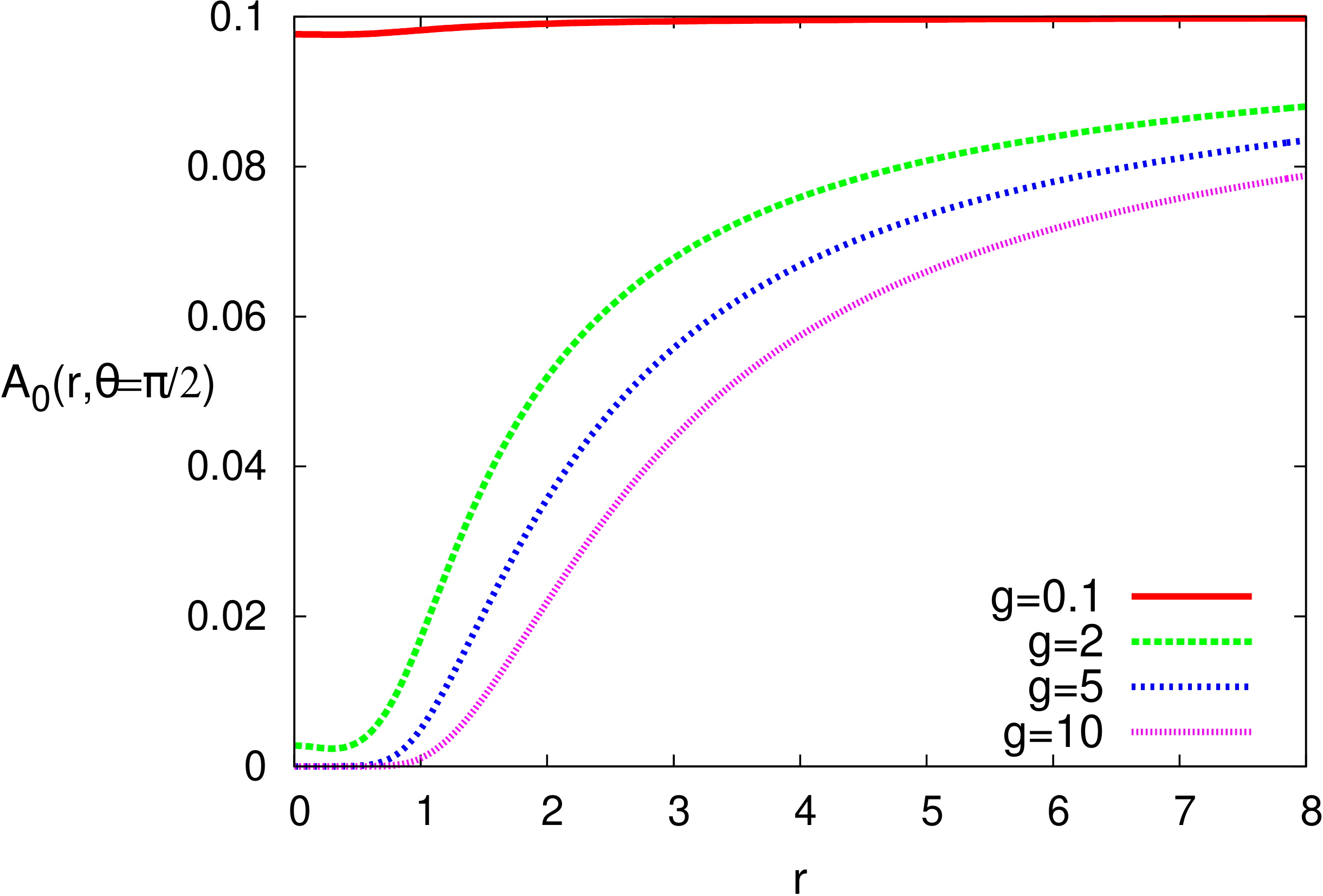}\qquad
\includegraphics[height=.215\textheight]{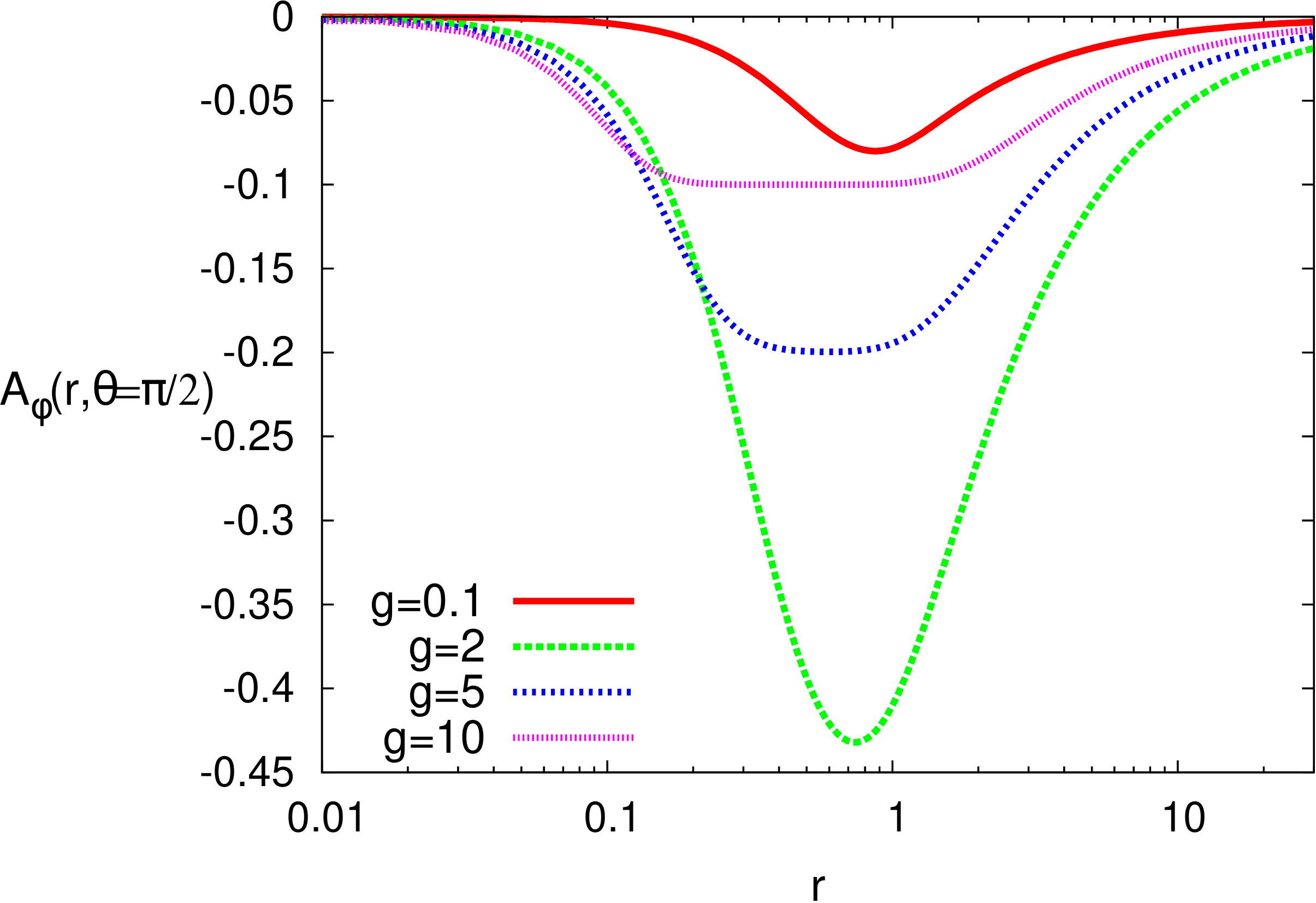}
\caption{Gauged $B=1$ Skyrmion: Profile functions of the component of the Skyrme field $\phi_0$, the electric potential $A_0$ and the magnetic potential~$A_\varphi$ of some illustrative solutions at the symmetry plane $\theta=\pi/2$ are plotted as functions of the radial coordinate for some set of values of $g$ at $V=0.1$.}\label{prof-g}
\end{figure}
The deformation of the Skyrmion is maximal as~$V$ becomes very small while the coupling constant~$g$ increases within allowed range, see Figure~\ref{eng-isosurfaces}. In such a limit, the electrostatic energy
 is at least two orders of magnitude smaller than the magnetic energy, as shown in Figure~\ref{eng-g} (right upper plot).
The total energy of
the gauged Skyrmion decreases as~$g$ increases,
 since the toroidal magnetic flux
squeezes the configuration towards the symmetry axis.

For relatively small values of the parameter $V$, the energy of electrostatic repulsion remains mach smaller that the magnetic energy of the gauged Skyrmion.
The Figure~\ref{prof-g} exhibits
the profiles in the $(xy)$-plane
of the $\phi_0$-component of the Skyrme field, the electrostatic potential~$A_0$ and of
the magnetic potential~$A_\varphi$,
for $V=0.1$ and several values of the gauge coupling constant~$g$.
Physically, the deformation of the Skyrmion is related with the
occurrence of the circular magnetic flux in the equatorial plane. Its position is associated with a minimum of the magnetic potential $A_\varphi$. As the gauge coupling increases within its allowed range, the electrostatic potential $A_0$
in some central region decreases,
approaching values close to zero for
large enough $g$,
see Figure~\ref{prof-g}, upper right plot.
\begin{figure}[h]\centering
\includegraphics[height=.22\textheight]{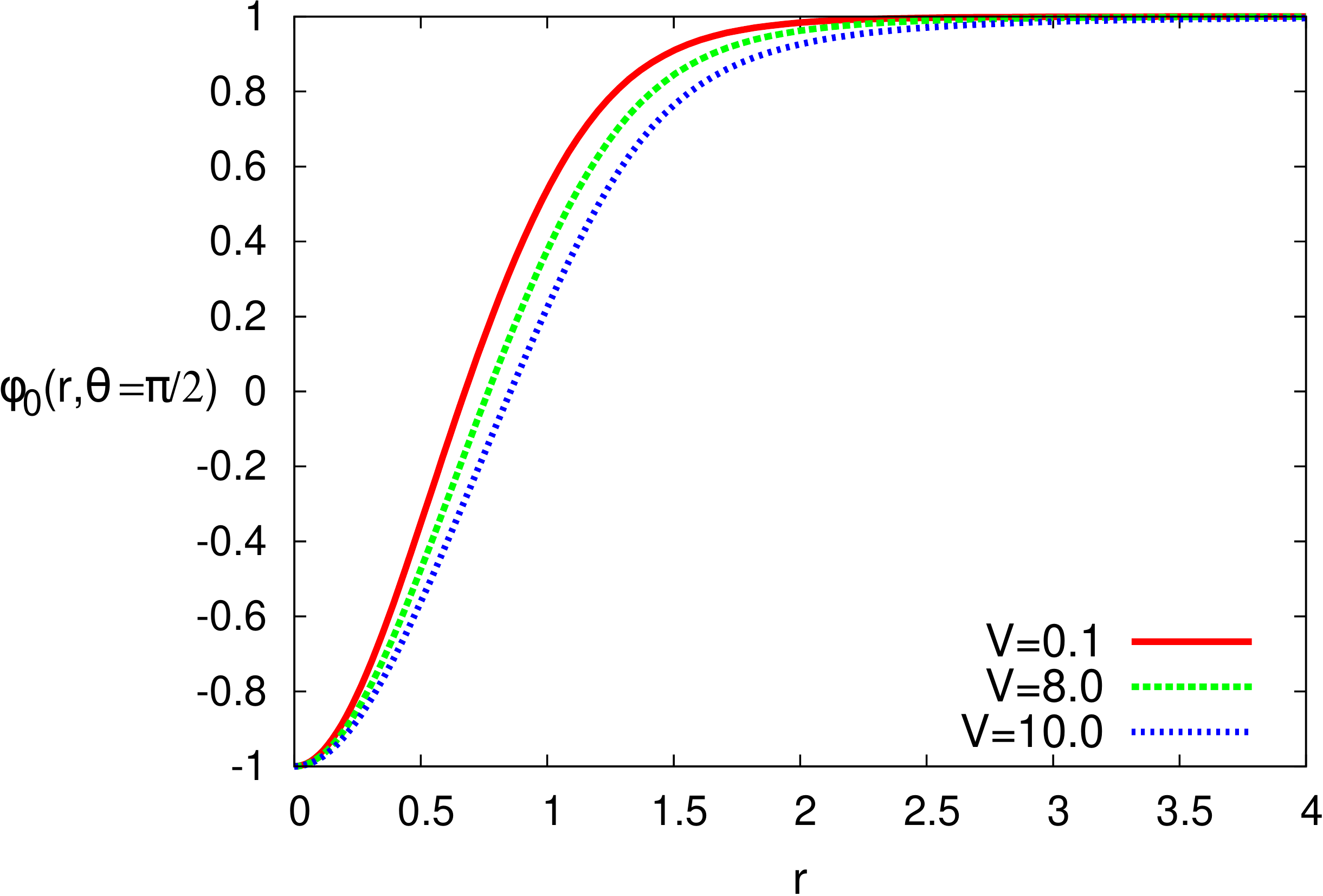} \
\includegraphics[height=.22\textheight]{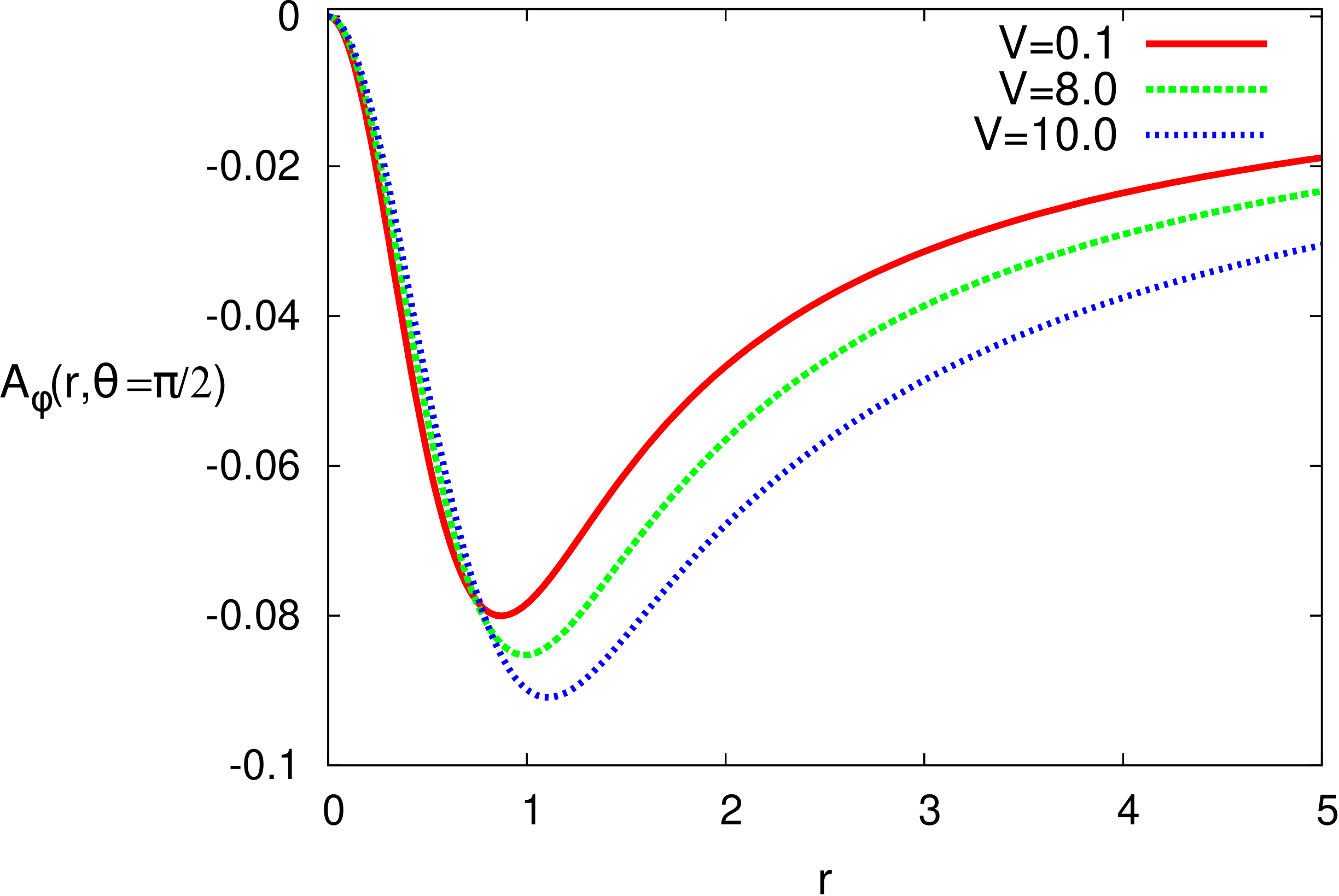}
\caption{Gauged $B=1$ Skyrmion: Profile functions of the component of the Skyrme field $\phi_0$ and the magnetic potential $A_\varphi$ of some illustrative solutions at the symmetry plane $\theta=\pi/2$ are plotted as functions of the radial coordinate for some set of values of $V$ at
$g=0.1$.}\label{prof-V}
\end{figure}

At the same time, in the strong coupling regime, the magnetic potential develops a sharp plateau in the equatorial plane, where $gA_\varphi+n \sim 0$ and
$A_0 \sim 0$, see Figure~\ref{prof-g}
(with $n$ the winding of the Skyrme field in the ansatz~\eqref{U}).
Clearly, within this domain,
the electromagnetic fields described by the Maxwell equations~\eqref{eqAphi},~\eqref{eqA0n} are massless.
Moreover, the electric field is pushed out of the core of the configurations,
in this limit the electric charge density distribution is almost spherically symmetric.
The plateau further extends to some region as the gauge coupling grows up to the maximal possible value. This pattern resembles
the Meissner effect: at some critical value of the gauge coupling $g_{\rm cr}\approx 3.3$ the magnetic field is expelled from the core and the circular magnetic flux becomes quantized in units of the topological winding~$n$.

 It is interesting to remark that
this situation is similar to that found
for the ${\rm U}(1)$ gauged Hopfions in the Faddeev--Skyrme--Maxwell model \cite{Shnir:2014mfa} and for the gauged planar Skyrmions \cite{Gladikowski:1995sc,Samoilenka:2015bsf,Shnir:2015twa}
and
gauged ${\rm O}(3)$ lumps~\cite{Schroers:1995he}.
In the limit $V=0$ the electric field is vanishing and the gauge coupling~$g$ may increase indefinitely. Effectively, this corresponds to the truncation of the system
to the case of the abelian Higgs model.
 Then the
linear string of magnetic flux generated by massless scalar excitations passes through
the center of the soliton core, which is strongly deformed by the ``supercurrent'' located in the equatorial plane.

\begin{figure}[th!]\centering
\includegraphics[height=.325\textheight, angle =-90]{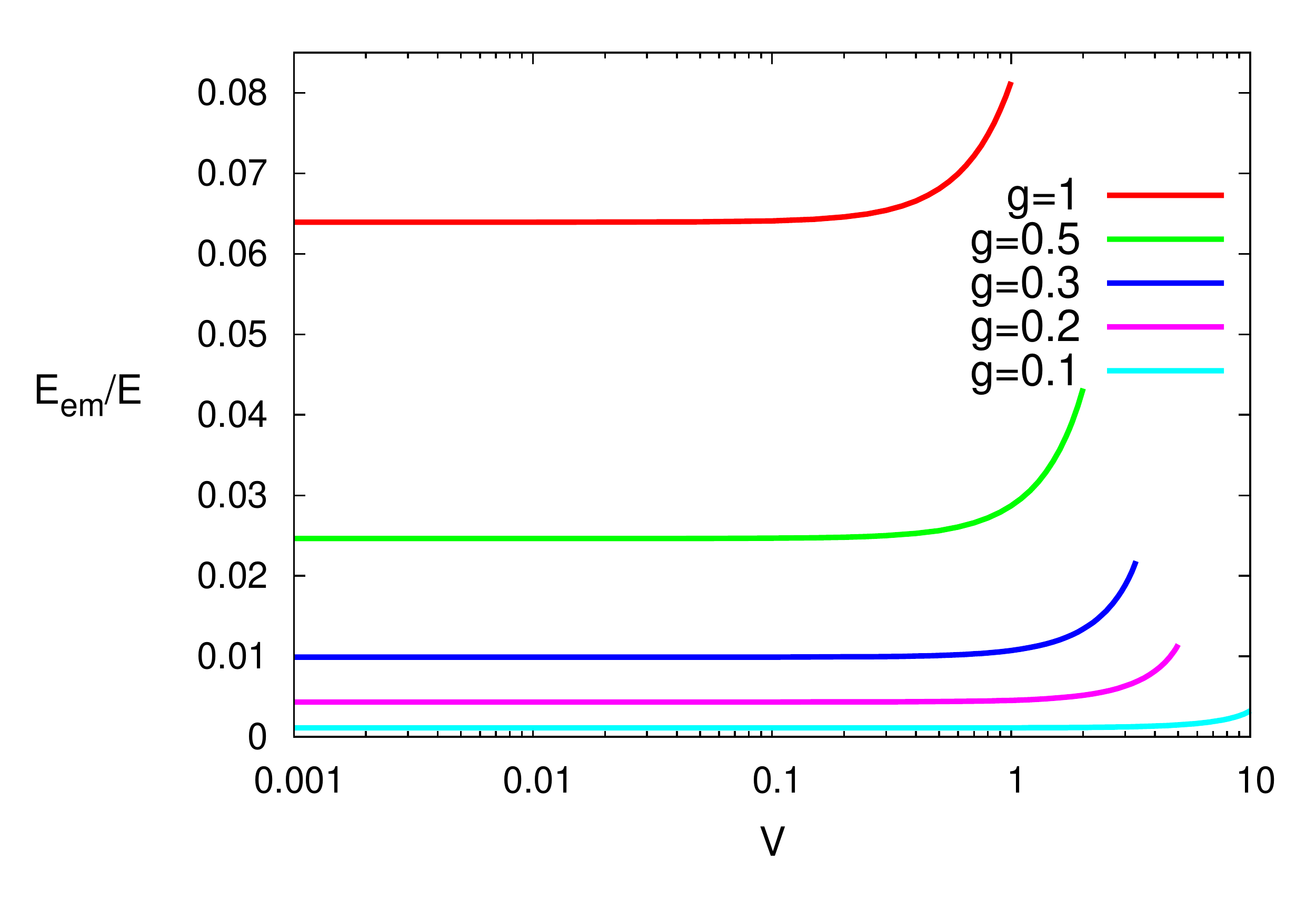}
\includegraphics[height=.325\textheight, angle =-90]{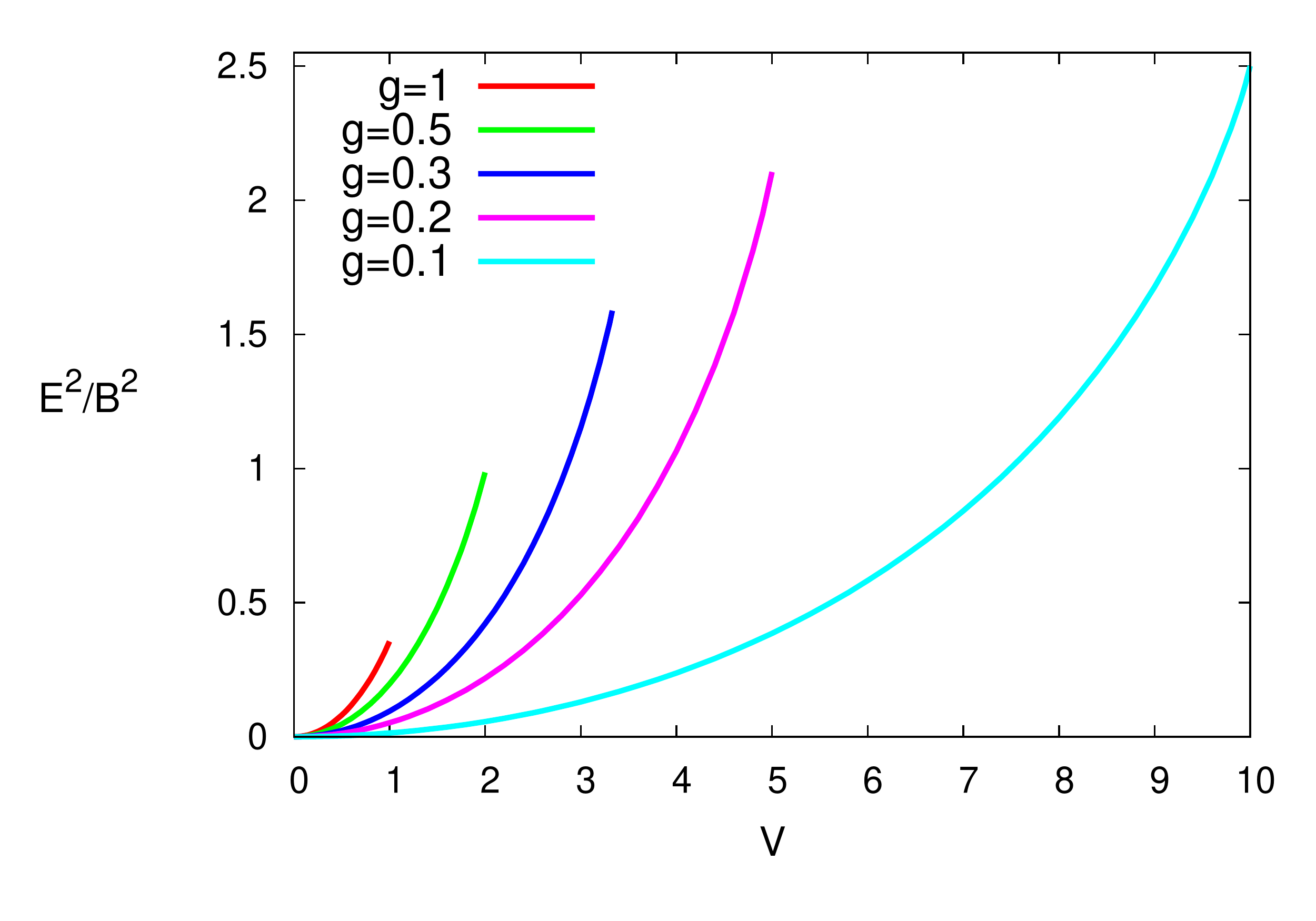}
\includegraphics[height=.325\textheight, angle =-90]{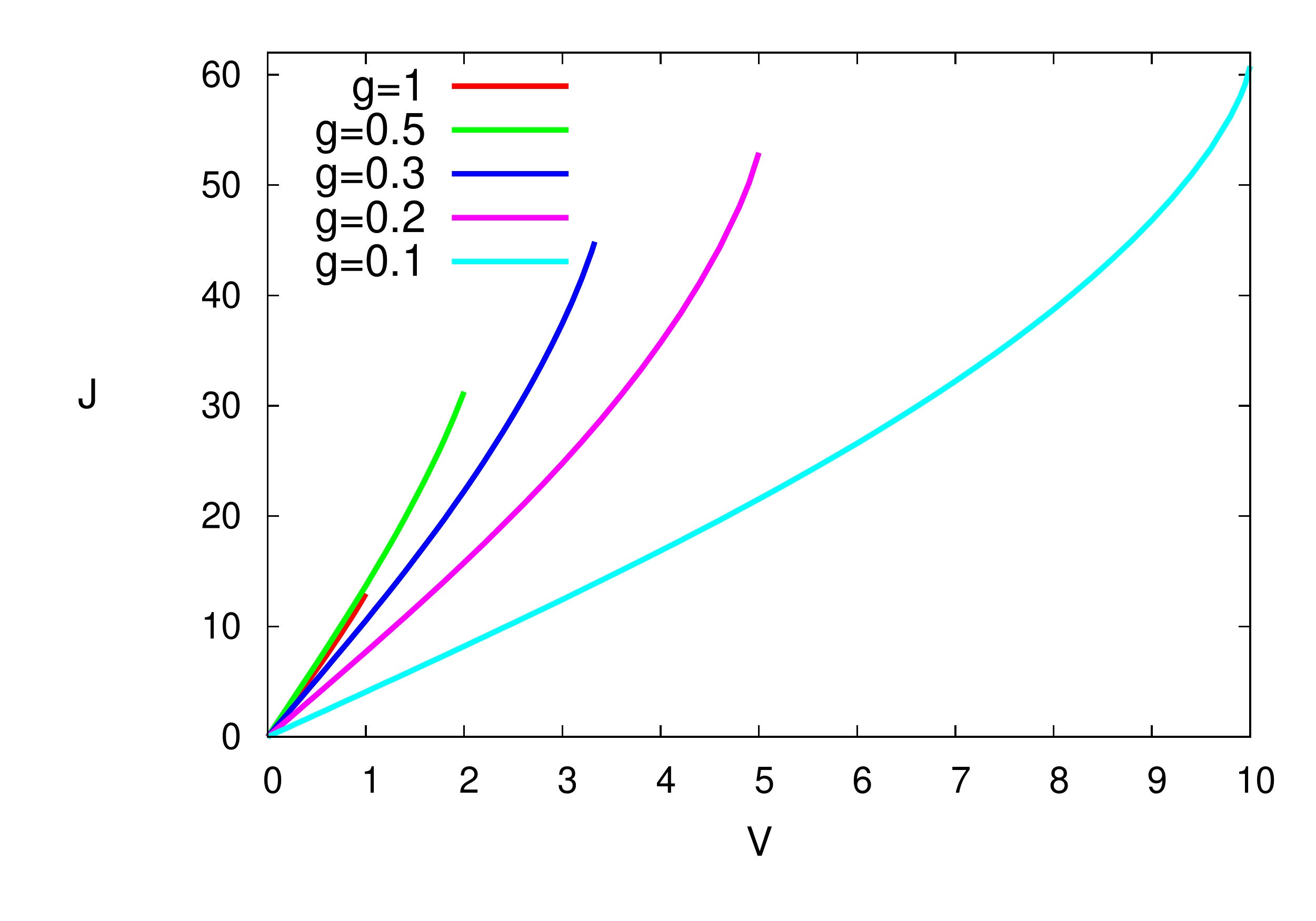}
\includegraphics[height=.325\textheight, angle =-90]{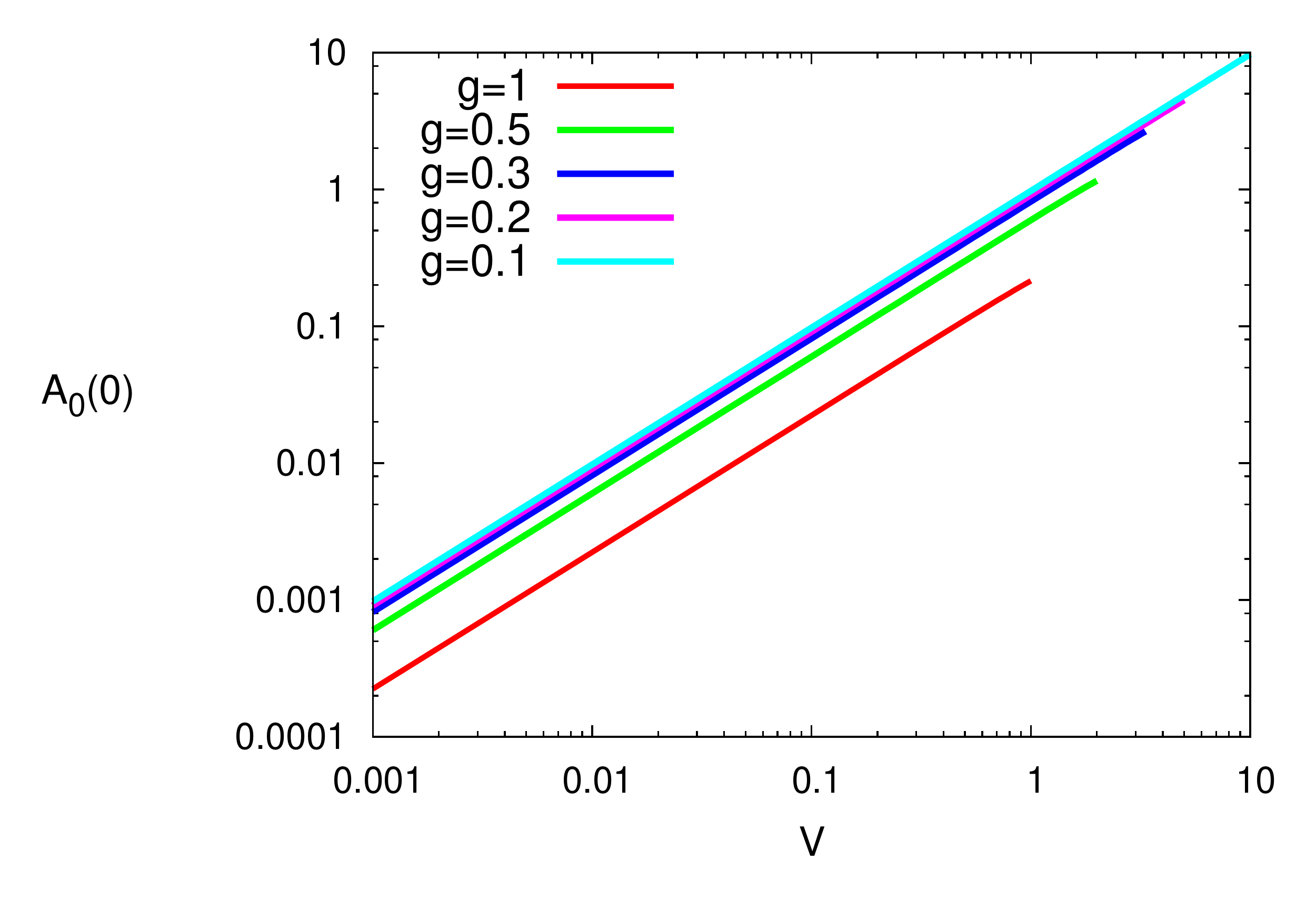}
\includegraphics[height=.325\textheight, angle =-90]{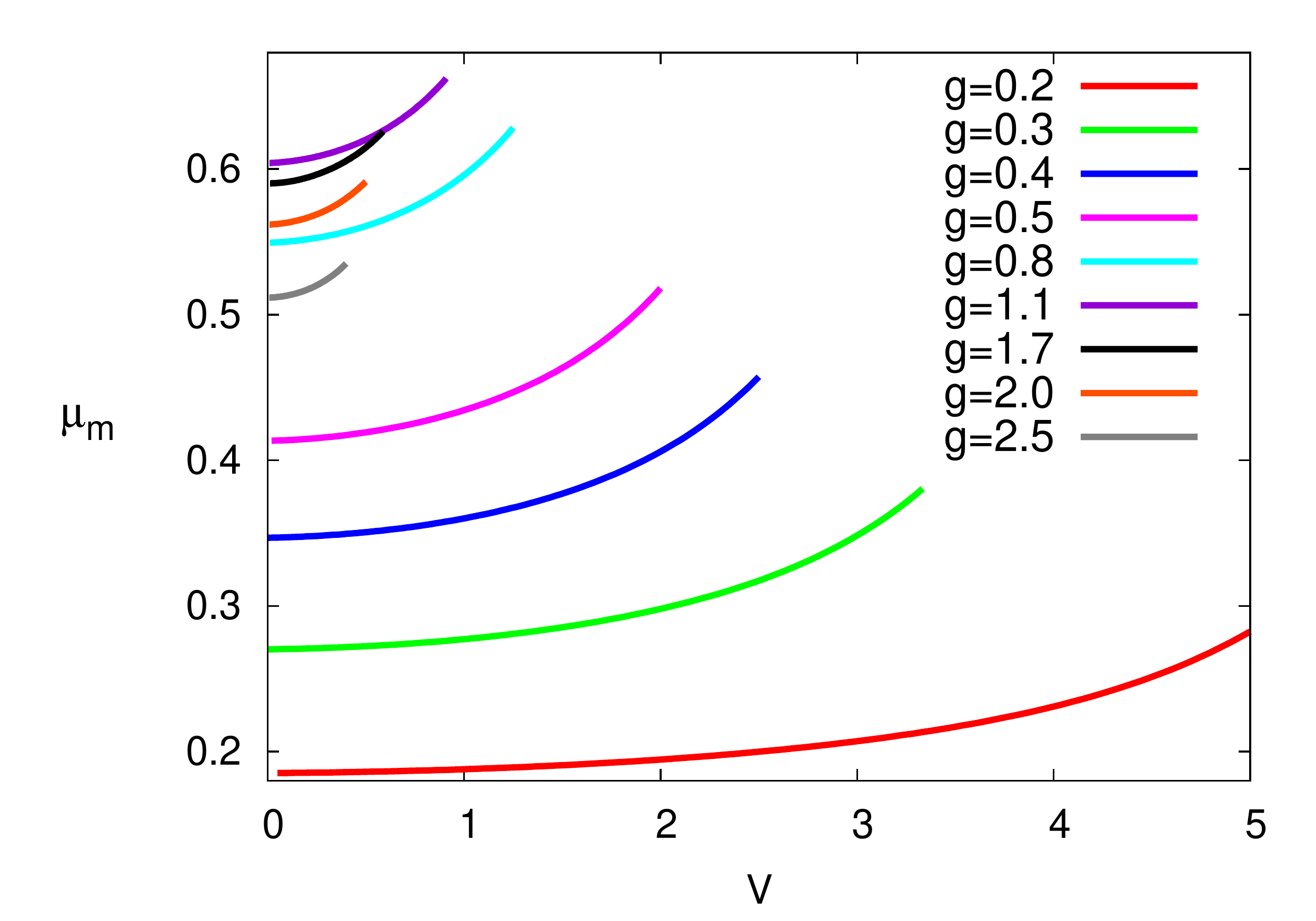}
\includegraphics[height=.325\textheight, angle =-90]{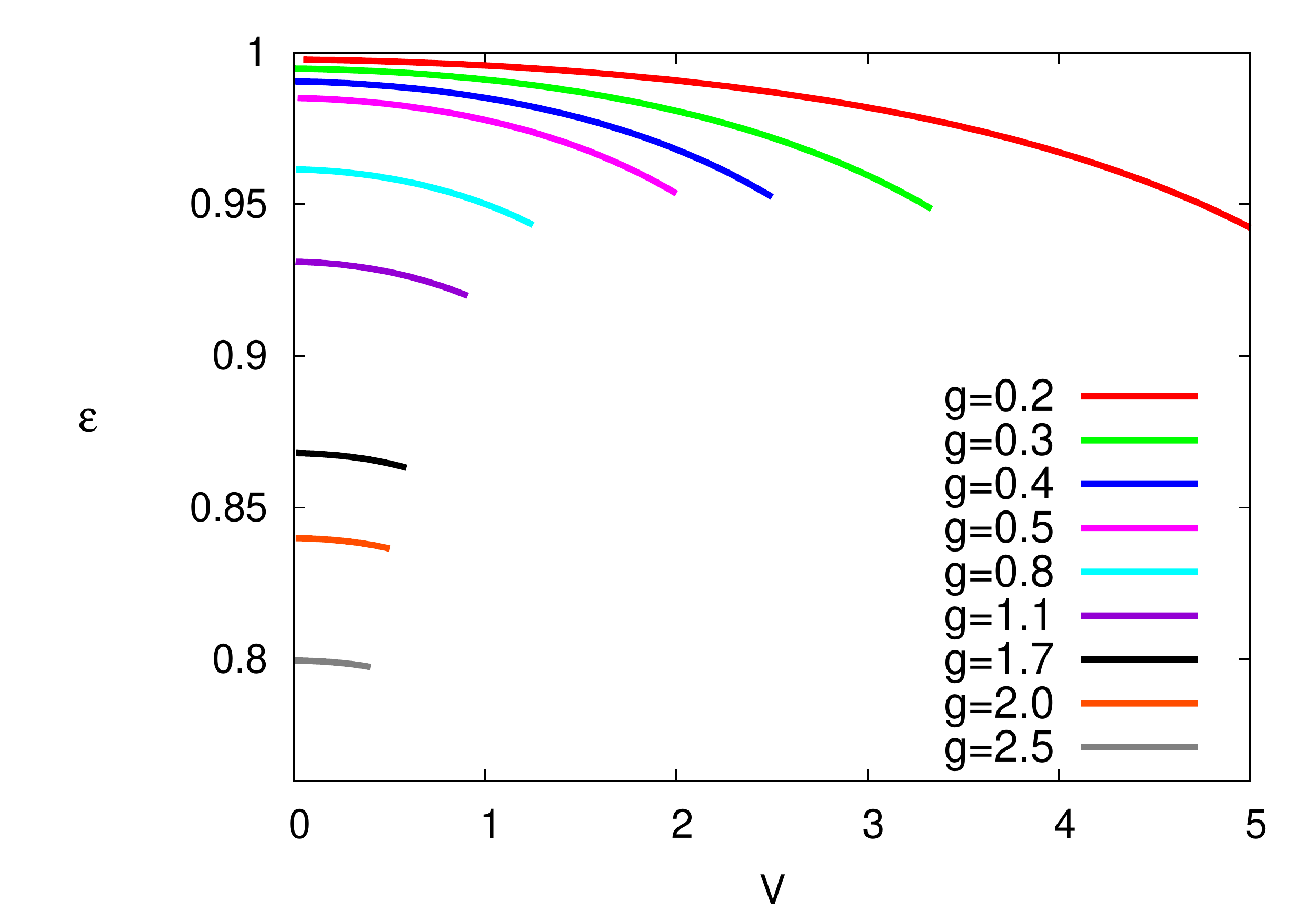}
\caption{Gauged $B=1$ Skyrmion: The ratio of the electromagnetic and total energies (upper left),
the ratio of the electric and magnetic energies (upper right), the total angular momentum (middle left),
the value of the electric potential $A_0$ at the center of the Skyrmion (middle right),
the values of the dipole magnetic moment
(bottom left) and the deformation parameter~$\epsilon$~\eqref{deform} (bottom right)
are plotted as
functions of the parameter $V$ for some set of values of the gauge coupling constant $g$ at $m=1$.}\label{eng-V}
\end{figure}

The electrostatic energy dominates when we set $g\ll1$ and increase the parameter $V$ up to its maximal possible value.
The angular momentum of the configuration rapidly increases, as displayed in Figure~\ref{eng-V}, middle left plot. The electrostatic repulsion
slightly increases the size of the core of the Skyrmion, see Figure~\ref{prof-V}. However, this effect
is not so strong, as it is in the case of dominance of the magnetic energy.

In Figure~\ref{eng-g}, we show the ratio between the electromagnetic and total energy of gauged
Skyrmion as defined by the functional \eqref{Heff},
as a function of the gauge coupling $g$ (upper left plot) for several values of $V$.
 For all considered range of parameters,
the contribution of electromagnetic energy to $E$ is less than 10 $\%$, this ratio being
maximal for $V\sim 0.5$ and $g\sim 2$.

 Also,
as the parameter $V$ remains smaller than $2$, the electrostatic energy of the configuration is smaller than the magnetic energy, see
Figure~\ref{eng-V}, upper right plot.
 However, the energy stored in the electric field
becomes more significant as $V$ increases and the gauge coupling $g$ remains relatively small.
For any $g$, both the magnetic dipole moment and the deformation of the solutions increase with $V$,
see Figure~\ref{eng-V} (bottom panels).
Moreover, note that for fixed $V$, the behaviour of $\mu_m$ as a function of $g$
 is non-monotonic.

Notably, both the energy and the angular momentum of gauged Skyrmion remain finite as $|g V| \to m$.
This is similar to the situation which was observed for isospinning solitons in $3+1$ dimensions
\cite{Battye:2014qva,Battye:2005nx,Harland:2013uk}.
On the contrary, in $2+1$ dimensions both the energy and the angular momentum of isospining Skyrmions diverge \cite{Battye:2013tka,Halavanau:2013vsa}.

Considering the dependency of the angular momentum and electric charge of the gauged Skyrmion, we observe that, for a fixed value of the electrostatic parameter~$V$,
both quantities become maximal at some value of the gauge coupling~$g$. Further increase of the coupling leads to decrease of the angular momentum~$J$, see Figure~\ref{eng-g}, lower left plot.

We can now attempt to apply our results to the physical properties of baryons. A standard approach is to make use of the Adkins--Nappi--Witten calibration setting the physical values of the Skyrme parameters as
$f_\pi =129$~MeV and $a_0=5.45$~\cite{Adkins:1983ya}. The pion mass parameter is taken as $m=0.526$, other calibrations are also suggested, see, e.g.,~\cite{Manton:2022}.

\begin{figure}[t]\centering
\includegraphics[height=.32\textheight, angle =-90]{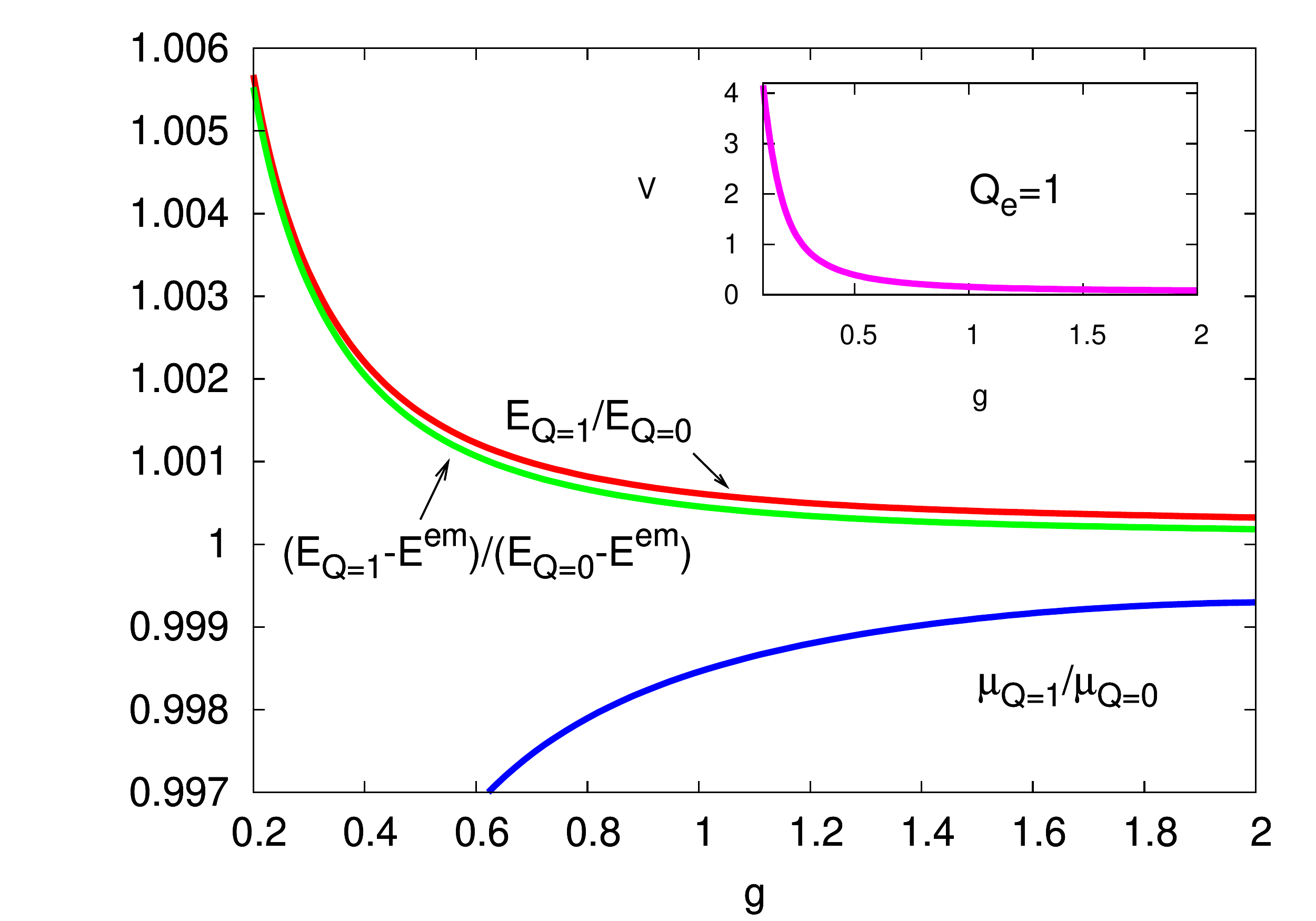}
\caption{Gauged $B=1$ Skyrmion: The ratio of energies of charged and uncharged Skyrmions and the ratio of the corresponding dipole magnetic moments are plotted as
functions of the gauge coupling $g$ at $m=0.526$. The curve of dependency $g(V)$ for $Q_e=1$ is displayed in the inset plot.}\label{pnsplit}\vspace{-1mm}
\end{figure}

Considering possible interpretation of gauged Skyrmions as baryons, we have to take into account that in natural units $\hbar =c=1$
the physical charge of a proton is $ Q_e^2=4\pi \alpha$ (see equation~\eqref{Qelectric}), where $\alpha=1/137$ is the fine structure constant. This condition yields a continuous set of values of parameters $(g,V)$, see unset plot it Figure~\ref{pnsplit}.

We have shown above that the gauged Skyrmion can lower its classical mass by interacting with the electromagnetic field. Since, naively, a neutron could be indentified with the ungauged $g=0$ Skyrmion, one can attempt to evaluate the corresponding mass splitting and compare it with corresponding experimental data.
However, real nuclei have non-zero magnetic dipole moments, therefore it will be consistent to consider a neutron as an electrically neutral gauged Skyrmion with $V=0$ and a~non-vanishing coupling~$g$.

Our numerical evaluations of the ratio of energies of charged and uncharged Skyrmions, which can be identified as a proton and a neutron, respectively, show that, for a given value of the gauge coupling $g$, the energy of a charged soliton is always higher, see Figure~\ref{pnsplit}. Clearly, this is not in a good agreement with the experimental data, the proton is a bit lighter than a~neutron. Subtraction of the contribution of electromagnetic energy does not change the situation, a proton and a neutron cannot be described in the gauged Skyrme model with the same set of parameters. Related evaluation of the ratio of magnetic moments of the $Q_e=1$ (in units of $\sqrt{4 \pi \alpha}$) and $Q_e=0$ states with fixed value of $g$ leads to an even worse disagreement, see Figure~\ref{pnsplit}. We can conclude that more accurate description of nuclei in the Skyrme model can be possible only via appropriate adjustment of the parameters of the model, even for the states with baryon charge~${B=1}$.

\section{Conclusions}\label{sec:5}

In this paper, we revisited the ${\rm U}(1)$ gauged Skyrme--Maxwell model
in $3+1$ dimensions
and studied the solutions of topological degree one.
We have shown that
their domain of existence is restricted by the value of the pion mass parameter~$m$,
 such that
 by analogy with the isospinning Skyrmions~\cite{Battye:2005nx},
the gauged Skyrmions do not exist in the model without a potential term. We confirm the observation of paper~\cite{Piette:1997ny} that the coupling to the electromagnetic field
violates the spherical symmetry of the configuration and induces a dipole magnetic moment of the Skyrmion, which carries both an electric charge and a (local)
magnetic flux (but not a net one).

We find that gauged Skyrmions exist for all
range of values of parameters of the model restricted by the condition $|g V| \le m$.
The upper critical value~$m$
yields two limiting cases, $|g| \gg |V|$ (magnetic limit) and $|V| \gg |g|$ (electrostatic limit), respectively.
The gauged Skyrmion is strongly deformed in the magnetic limit, it becomes extremely elongated and stretched out by the circular magnetic flux. In the opposite limit the repulsive electrostatic interaction increases the size of the interior region of the soliton, however the deformation of the Skyrmion if not so strong as in the magnetic limit.

Our results show that, although the mass of a
gauged Skyrmion can be decreased by
interaction with the electromagnetic field, the
proton-neutron mass difference cannot be described in the ${\rm U}(1)$ gauged Skyrme model with the same fixed set of parameters.

Certainly, this is a first step toward complete investigation
of the gauged Skyrmions, this study should be extended to the solutions of higher degrees and different
geometry. One can expect the electrostatic repulsion and generation of magnetic fluxes may significantly affect the bounded configurations of gauged Skyrmions.

\vspace{-1mm}

\subsection*{Acknowledgements}
We are grateful to D.H.~Tchrakian for inspiring and valuable discussions.
 The work of E.R.\ is supported by the Center for Research and Development in Mathematics and Applications (CIDMA) through the Portuguese Foundation for Science and Technology (FCT -- Fundac\~ao para a Ci\^encia e a Tecnologia), references UIDB/04106/2020 and UIDP/04106/2020.
E.R.\ also acknowledge support from the projects CERN/FIS-PAR/0027/2019, PTDC/FIS-AST/3041/2020, CERN/FIS-PAR/0024/2021 and 2022.04560.PTDC.
This work has further been supported by the European Union's Horizon 2020 research and innovation (RISE) programme H2020-MSCA-RISE-2017 Grant No.~FunFiCO-777740 and by the European Horizon Europe staff exchange (SE) programme HORIZON-MSCA-2021-SE-01 Grant No.~NewFunFiCO-101086251.

\pdfbookmark[1]{References}{ref}
\LastPageEnding

\end{document}